\documentclass[10pt,fullpage,onecolumn,a4paper]{article}
\title {New Probabilistic Interest Measures for Association Rules}
\author {Michael Hahsler and Kurt Hornik\\ 
Vienna University of Economics and Business Administration,\\
Augasse 2--6, A-1090 Vienna, Austria.
}
\usepackage{amsmath}
\usepackage{amsfonts}
\usepackage{latexsym}
\usepackage{graphicx}
\usepackage{hyperref}
\usepackage[numbers]{natbib}
\begin{document}
\maketitle
%
%
%
\begin{abstract}
  Mining association rules is an important technique for discovering
  meaningful patterns in transaction databases. Many different measures
  of interestingness have been proposed for association rules. However,
  these measures fail to take the probabilistic properties of the mined
  data into account. We start this paper with presenting a simple
  probabilistic framework for transaction data which can be used to
  simulate transaction data when no associations are present. We use
  such data and a real-world database from a grocery outlet to explore
  the behavior of confidence and lift, two popular interest measures
  used for rule mining. The results show that confidence is
  systematically influenced by the frequency of the items in the left
  hand side of rules and that lift performs poorly to filter random
  noise in transaction data. Based on the probabilistic framework we
  develop two new interest measures, hyper-lift and hyper-confidence,
  which can be used to filter or order mined association rules.  The new
  measures show significantly better performance than lift for
  applications where spurious rules are problematic.

\end{abstract}



\section{Introduction}
Mining association rules~\citep{arules:Agrawal1993}
is an important technique for discovering
meaningful patterns in transaction databases.
An association rule is a rule of the form $X \Rightarrow Y$,
where $X$ and $Y$ are two disjoint sets of items (itemsets).
The rule means that if we find all items in~$X$ in a transaction
it is likely that the transaction also contains the items in~$Y$.


Association rules are selected from the set of all possible rules
using measures of significance and interestingness.
\emph{Support}, the primary measure of significance, is defined as
the fraction of transactions in the database which contain all items in 
a specific rule~\citep{arules:Agrawal1993}. That is,
\begin{equation}
\mathrm{supp}(X \Rightarrow Y) = \mathrm{supp}(X \cup Y) = 
\frac{c_{XY}}{m},
\end{equation}
where $c_{XY}$ represents the number of transactions
which contain all items in $X$ and $Y$, and $m$ is the number of 
transactions in the database.

For association rules, a minimum support threshold is used to select the most
frequent (and hopefully important) item combinations called \emph{frequent
itemsets}.  The process of finding these frequent itemsets in a large database
is computationally very expensive since it involves searching a lattice which,
in the worst case, grows exponentially in the number of items. In the last
decade, research has centered on solving this problem and a variety of
algorithms were introduced which render search feasible by exploiting various
properties of the lattice (see~\cite{arules:Goethals2004} for pointers to the
currently fastest algorithms).

From the frequent itemsets all rules which satisfy a threshold on a
certain measures of interestingness are generated.  For association rules,
\citet{arules:Agrawal1993} suggest using a threshold on \emph{confidence}, one
of many proposed measures of interestingness.  A practical problem is that with
support and confidence often too many association rules are produced.  One
possible solution is to use additional interest measures, such as
e.g.~\emph{lift} \citep{arules:Brin1997}, to further filter or rank found rules. 

Several authors \cite{arules:Brin1997, arules:Aggarwal1998,
  arules:Silverstein1998, arules:Adamo2001} constructed examples to show
that in some cases the use of confidence and lift can be problematic.
Here, we instead take a look at how pronounced and how important such
problems are when mining association rules. To do this, we visually
compare the behavior of support, confidence and lift on a transaction
database from a grocery outlet with a simulated data set which only
contain random noise. The data set is simulated using a simple
probabilistic framework for transaction data (first presented
by~\citet{hahsler:Hahsler2006b}) which is based on independent Bernoulli trials
and represents a null model with ``no structure.''

Based on the probabilistic approach used in the framework, we will develop 
and analyze two new measures of interestingness, \emph{hyper-lift} and
\emph{hyper-confidence.} We will show how these measures are better suited to 
deal with random noise and that the measures do not suffer from the
problems of confidence and lift.
 
This paper is structured as follows: In Section~\ref{sec:modeling}, we
introduce the probabilistic framework for transaction data.  In
Section~\ref{sec:data}, we apply the framework to simulate a comparable data
set which is free of associations and compare the behavior of the measures
confidence and lift on the original and the simulated data.  Two new interest
measures are developed in Section~\ref{sec:new_measures} and compared on three
different data sets with lift.  We conclude the paper with the main findings
and a discussion of directions for further research.

An implementation of the probabilistic framework and the new measures of
interestingness proposed in this paper is included in the freely available
R~extension package
\textbf{arules}~\citep{arules:Hahsler+Grun+Hornik:package:2006}\footnote{R is a
free software environment for statistical computation, data analysis and
graphics. The R software and the extension package \textbf{arules} are
available for download from the Comprehensive R Archive Network (CRAN) under
\url{http://CRAN.R-project.org/}.}.

\section{A simple probabilistic framework for transaction data}
\label{sec:modeling}

\begin{figure}[tp]
\centering
\includegraphics[width=7cm]{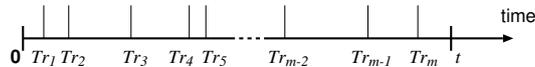}
\caption{Transactions occurring over time following a 
  Poisson process.\label{fig:transactions}}
\end{figure}

A transaction database consists of a series of transactions, each transaction
containing a subset of the available items.  We consider transactions which are
recorded during a fixed time interval of length~$t$.  In
Figure~\ref{fig:transactions} an example time interval is shown as an arrow
with markings at the points in time when the transactions denoted by
$\mathit{Tr}_1$ to $\mathit{Tr}_m$ occur.  For the model we assume that
transactions occur randomly following a (homogeneous) Poisson process with
parameter~$\theta$.  The number of transactions~$m$ in time interval~$t$ is
then Poisson distributed with parameter~$\theta t$ where $\theta$ is the
intensity with which transactions occur during the observed time interval:
\begin{equation}
P(M=m) = \frac{e^{-\theta t} (\theta t)^m}{m!}
\label{equ:poisson}
\end{equation}

We denote the items which occur in the database by
$L=\{l_1,l_2,\ldots,l_n\}$ with $n$ being the number of different items.
For the simple framework we assume that all items occur independently of each
other and that for each item $l_i \in L$ there exists a fixed
probability~$p_i$ of being contained in a transaction.  Each transaction
is then the result of $n$ independent Bernoulli trials, one for each
item with success probabilities given by the
vector~$p=(p_1,p_2,\dots,p_n)$.  Table~\ref{tab:database} contains the
typical representation of an example database as a binary incidence
matrix with one column for each item.  Each row labeled $\mathit{Tr}_1$
to $\mathit{Tr}_m$ contains a transaction, where a 1 indicates presence
and a 0 indicates absence of the corresponding item in the transaction.
Additionally, in Table~\ref{tab:database} 
the success probability for each item is given in
the row labeled~$p$ and the row labeled~$c$ contains the number
of transactions each item is contained in (sum of the ones per column).

\begin{table}[tp]
\centering
\begin{tabular}{cccccc}
  & \multicolumn{5}{c}{items} \\
transactions & $l_1$ & $l_2$ & $l_3$ & \ldots & $l_n$ \\
\hline
$Tr_1$ & $0$ & $1$ & $0$ & \dots & $1$ \\
$Tr_2$ & $0$ & $1$ & $0$ & \dots & $1$ \\
$Tr_3$ & $0$ & $1$ & $0$ & \dots & $0$ \\
$Tr_4$ & $0$ & $0$ & $0$ & \dots & $0$ \\
$\vdots$ & $\vdots$ & $\vdots$ & $\vdots$ & $\ddots$ & $\vdots$ \\
$Tr_{m-1}$ & $1$ & $0$ & $0$ & \dots & $1$ \\
$Tr_{m}$ & $0$ & $0$ & $1$ & \dots & $1$ \\
\hline
$c$ & $99$ & $201$ & $7$ & \dots & $411$ \\
\hline
$p$ & $0.005$ & $0.01$ & $0.0003$ & \dots & $0.025$ \\

\end{tabular}
\caption{Example transaction database with 
transaction counts per item $c$
and items success probabilities $p$.\label{tab:database}}
\end{table}

Following the model, $c_i$, the observed number of transactions
item~$l_i$ is contained in, can be interpreted as a realization of a
random variable~$C_i$.  Under the condition of a fixed number of
transactions~$m$, this random variable has the following binomial
distribution.

\begin{equation}
P(C_i=c_i | M=m) = \binom{m}{c_i} p_i^{c_i} (1-p_i)^{m-c_i}
\end{equation}

However, since for a given time interval the number of transactions is
not fixed, the unconditional distribution gives:
\begin{equation}
\begin{split}
P(C_i=c_i) &= \sum_{m=c_i}^\infty P(C_i=c_i | M=m) \cdot P(M=m) \\
     &= \sum_{m=c_i}^\infty \binom{m}{c_i} p_i^{c_i} (1-p_i)^{m-c_i}\;
	\frac{e^{-\theta t} (\theta t)^m}{m!}        \\
     &= \frac{e^{-\theta t} (p_i \theta t)^{c_i}}{c_{i}!}
     	\sum_{m=c_i}^\infty \frac{((1-p_i)\theta t)^{m-c_i}}{(m-c_i)!} \\
     &= \frac{e^{-p_i \theta t} (p_i \theta t)^{c_i}}{c_i!}.  \\
\label{eq:prob}
\end{split}
\end{equation}

The term $\sum_{m=c_i}^\infty \frac{((1-p_i)\theta t)^{m-c_i}}{(m-c_i)!}$ in
the second to last line in Equation~\ref{eq:prob} is an exponential series with
sum $e^{(1-p_i)\theta t}$.  After substitution we see that the unconditional
probability distribution of each $C_i$ follows a Poisson distribution with
parameter~$p_i \theta t$.  For short we will use $\lambda_i = p_i \theta t$ and
introduce the parameter vector $\lambda=(\lambda_1,\lambda_2,\dots,\lambda_n)$
of the Poisson distributions for all items.  This parameter vector can be
calculated from the success probability vector~$p$ and vice versa by the linear
relationship $\lambda = p \theta t$.

For a given database, the values of the parameter $\theta$ and the
success vectors $p$ or alternatively $\lambda$ are unknown but can be
estimated from the database.  The best estimate for $\theta$ from a
single database is $m / t$.  The simplest estimate for $\lambda$ is to
use the observed counts $c_i$ for each item.  However, this is only a very
rough estimate which gets especially unreliable for small counts.  There
exist more sophisticated estimation approaches.  For example,
\citet{arules:DuMouchel2001} use the assumption that the parameters of the count
processes for items in a database are distributed according to a
continuous parametric density function.  This additional information can
improve estimates over using just the observed counts.  

Alternatively, the parameter vector $p$ can be drawn from a 
parametric distribution. A suitable distribution is the Gamma distribution 
which is very flexible and allows to fit a wide range of empirical data. 
A Gamma distribution together with the independence model introduced above 
is known as the Poisson-Gamma mixture model
which results in a negative binomial distribution
and has applications in many fields~\citep{Johnson1993}.
In conjunction with association rules this mixture model was used 
by \citet{hahsler:Hahsler2006a} to develop a model-based 
support constraint.

Independence models similar to the probabilistic framework
employed in this paper have been used for other applications.
In the context of query approximation,
where the aim is to predict the results of a query without
scanning the whole database,
\citet{arules:Pavlov2003}
investigated the independence model as an extremely parsimonious model.
However, the quality of the approximation can be poor if the 
independence assumption is violated significantly by the data.

\citet{arules:Cadez2001} and
\citet{arules:Hollmen2003} used the independence model to
cluster transaction data by learning the 
components of a mixture of independence models.
In the former paper the aim is to identify typical customer profiles 
from market basket data for outlier detection, customer ranking and 
visualization.
The later paper focuses on approximating the joint probability distribution
over all items by mining frequent itemsets in each component of the mixture
model, using the maximum entropy technique to obtain local models for
the components, and then combining the local models.

Almost all authors use the independence model to learn something from the
data. However, the model only uses the marginal probabilities $p$ of the
items and ignores all interactions. Therefore, the accuracy and usefulness
of the independence model for such applications is drastically limited and
models which incorporate pair-wise or even higher interactions provide better 
results.
For the application in this paper, we explicitly want to generate data 
with independent items to evaluate measures of interestingness.

%
%
\section{Simulated and real-world database}
\label{sec:data}

We use 1 month ($t=30$~days) of real-world point-of-sale transaction
data from a typical local grocery outlet.  For convenience reasons we
use categories (e.g., \emph{popcorn}) instead of the individual brands.
In the available $m=9835$ transactions we found $n=169$ different
categories for which articles were purchased.  This database is called
``Grocery'' and is freely distributed with the R~extension package
\textbf{arules}~\citep{arules:Hahsler+Grun+Hornik:package:2006}.

The estimated transaction intensity $\theta$ for Grocery 
is $m/t = 327.5$ transactions per day.
To simulate comparable data using the framework,
we use the Poisson distribution with the parameter
$\theta t$ to draw the number of transactions~$m$ 
(9715 in this experiment). 
For simplicity we use the relative observed item frequencies as estimates
for $\lambda$ and calculate the 
success probability vector~$p$ by $\lambda / \theta t$. 
With this information we simulate the $m$ transactions in the 
transaction database.
Note, that the simulated database does not contain any associations (all items
are independent), and thus differs from the Grocery database
which is expected to contain associations.
In the following we will use the simulated data set 
not to compare it to the real-world data set, but to
show that interest measures used for association rules
exhibit similar effects on real-world data as on
simulated data without any associations.

For the rest of this section we concentrate on 
$2$-itemsets, i.e., the co-occurrences between two items denoted by
$l_i$ and $l_j$ with $i,j = 1,2,\dots,n$ and
$i \ne j$.  
Although itemsets and rules of arbitrary length can be analyzed using 
the framework,
we restrict the analysis to $2$-itemsets
since interest measures for these associations are easily visualized 
using 3D~plots.
In these plots the $x$ and $y$-axis each represent the items 
$l_i$ and $l_j$
ordered
from the most frequent to the least frequent from left to right 
and front to back. On 
the $z$-axis we plot the analyzed measure.

\begin{figure}[pt]
\begin{minipage}[b]{.50\linewidth}
\centering
\includegraphics[width=\linewidth]{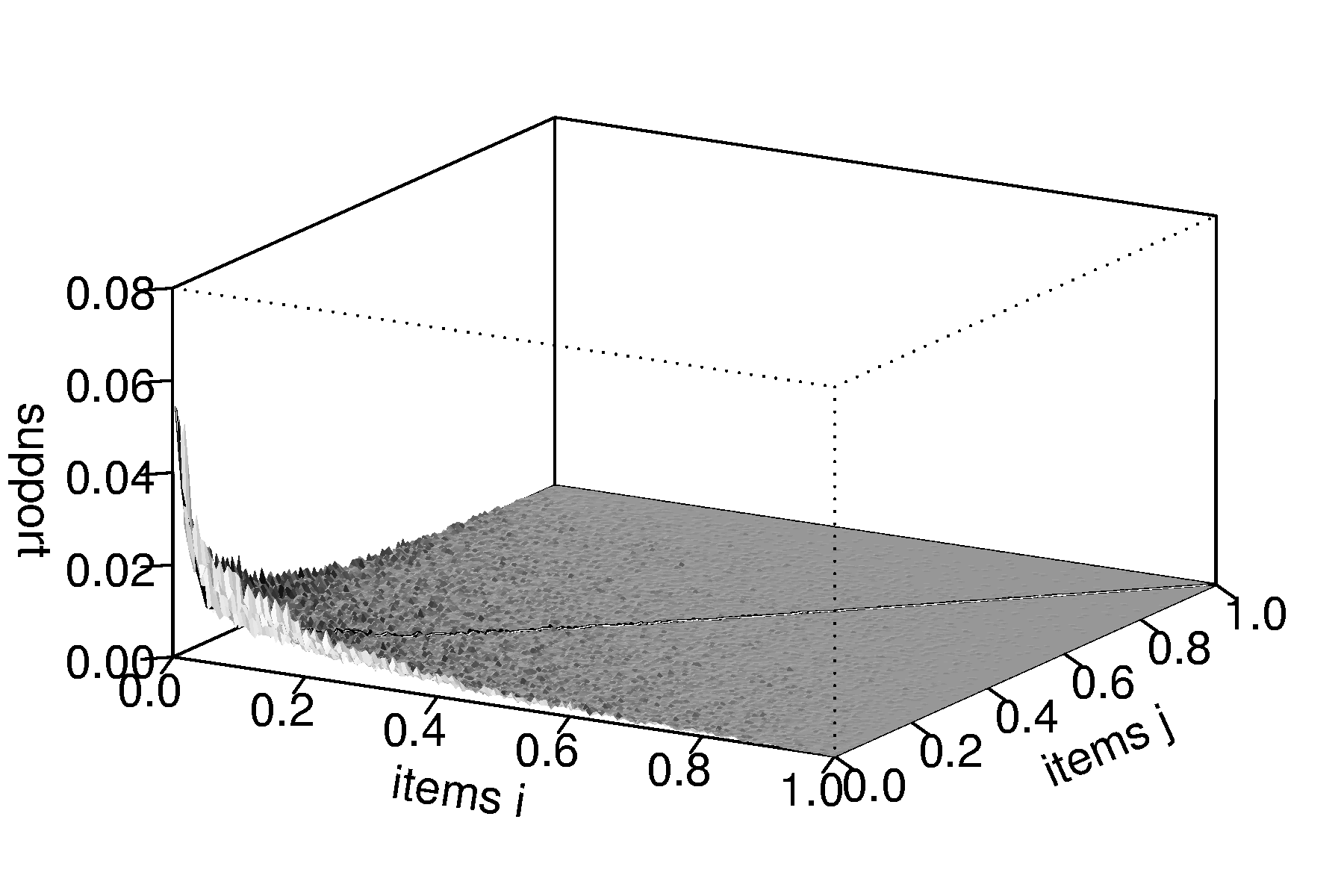}
(a) simulated
\end{minipage}
\begin{minipage}[b]{.50\linewidth}
\centering
\includegraphics[width=\linewidth]{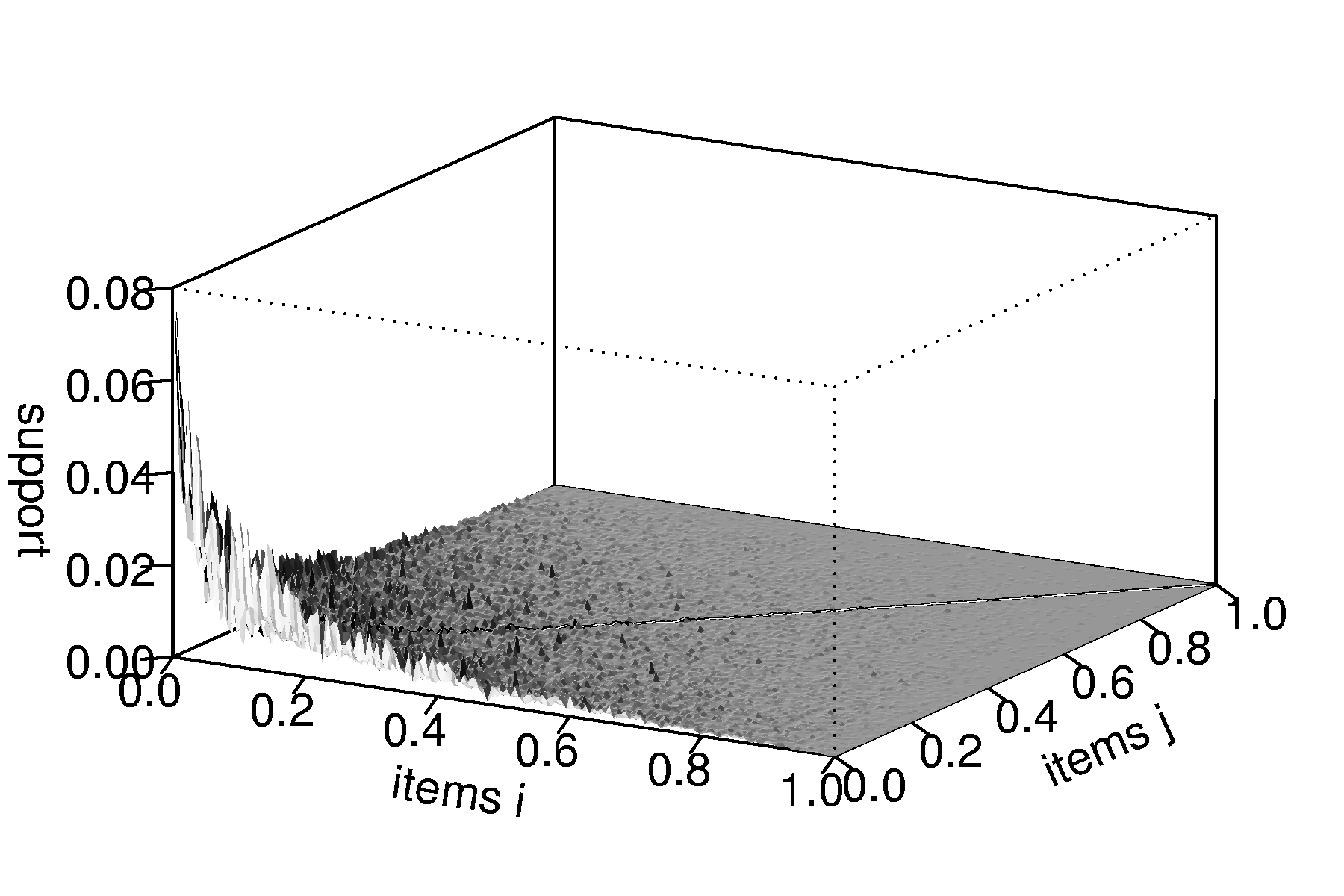}
(b) Grocery
\end{minipage}
\caption{Support distributions of all $2$-itemsets 
(items are ordered by decreasing support from left to right and front to back).\label{fig:supp}}
\end{figure}

First we compare the $2$-itemset support. 
Figure~\ref{fig:supp} 
shows the support distribution of all $2$-itemsets. 
Naturally, the most frequent items also form together 
the most frequent itemsets
(to the left in the front of the plots).
The general forms of the two support distributions 
in the plot are very similar.
The Grocery data set reaches higher support values
with a median of $0.000203$ 
compared to $0.000113$
for the simulated data. This indicates that the Grocery data set 
contains associated items which co-occur more often than expected 
under independence.

\subsection{The interest measure confidence}
\label{sec:conf}

Confidence is defined by \citet{arules:Agrawal1993} as
\begin{equation}
\mathrm{conf}(X \Rightarrow Y) =
\frac{\mathrm{supp}(X \cup Y)}{\mathrm{supp}(X)},
\label{equ:conf}
\end{equation}
where $X$ and $Y$ are two disjoint itemsets.  Often confidence is
understood as an estimate of the conditional probability 
$P(E_Y|E_X)$,
were $E_X$ ($E_Y$) is the event that $X$ ($Y$) occurs in a
transaction~\cite{arules:Hipp2000}.

\begin{figure}[pt]
\begin{minipage}[b]{.50\linewidth}
\centering
\includegraphics[width=\linewidth]{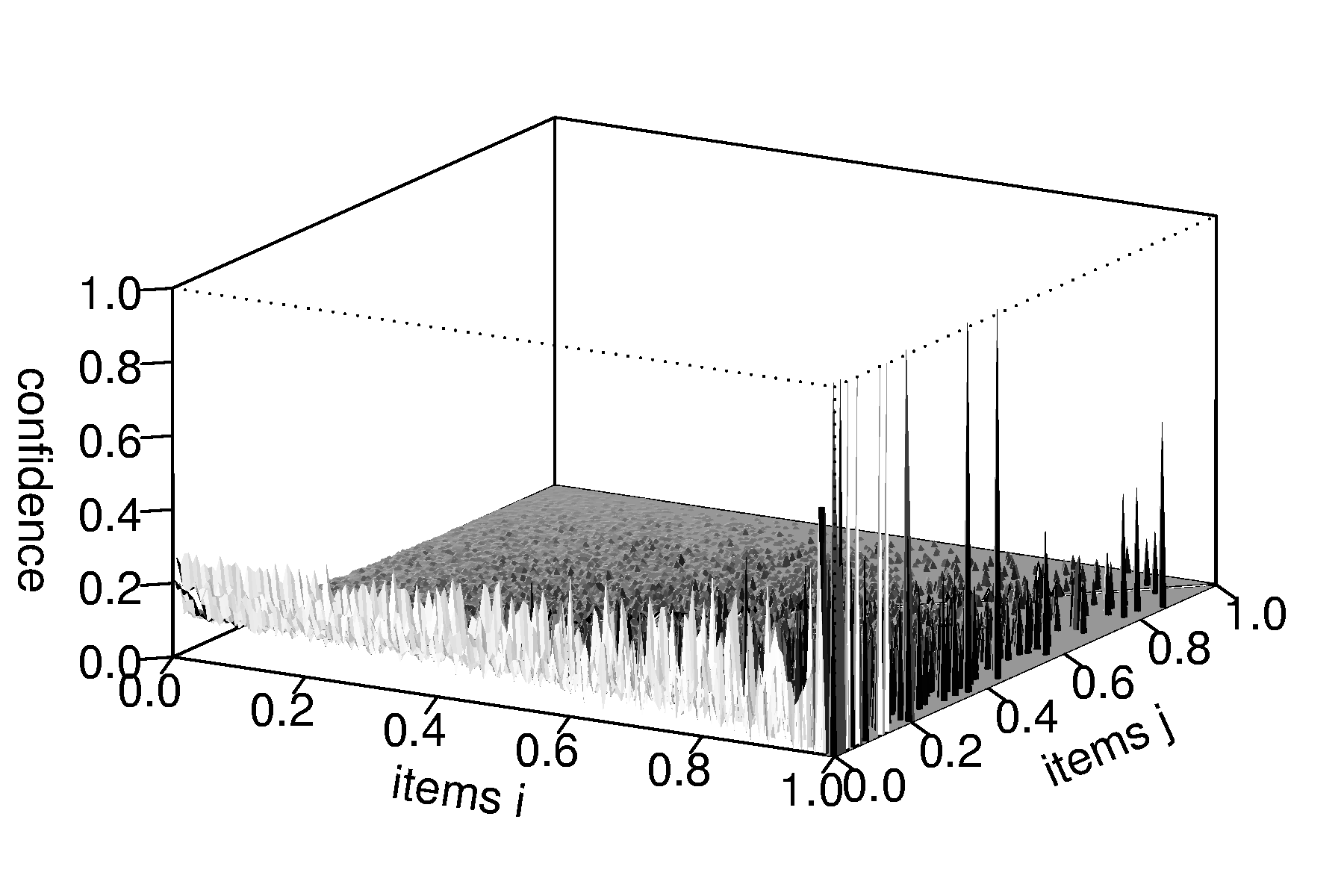}
(a) simulated
\end{minipage}
\begin{minipage}[b]{.50\linewidth}
\centering
\includegraphics[width=\linewidth]{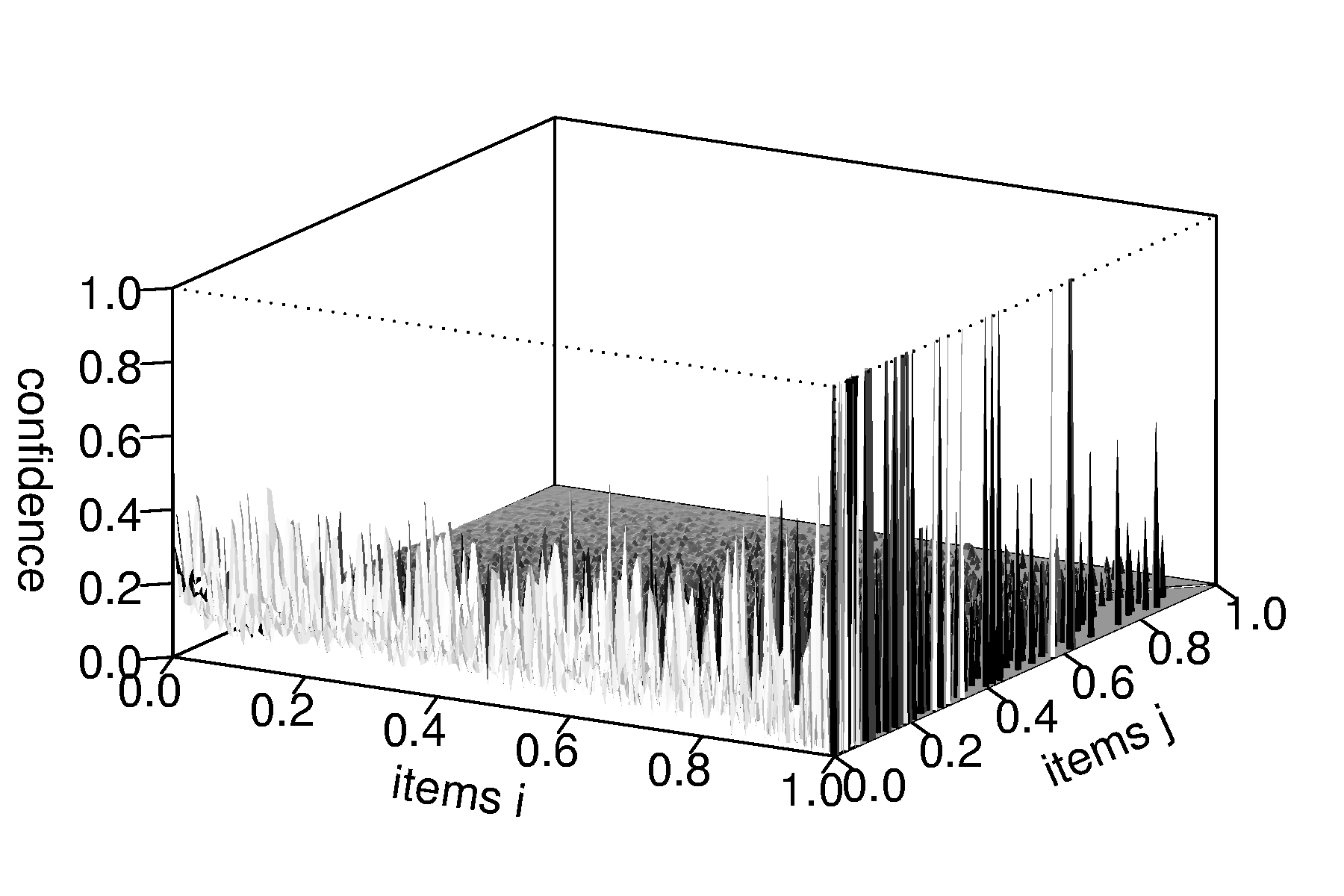}
(b) Grocery
\end{minipage}
\caption{Confidence distributions of all rules containing 2 items.\label{fig:conf}}
\end{figure}

From the $2$-itemsets we generate all rules of the from $l_i \Rightarrow l_j$
and present the confidence distributions in Figures~\ref{fig:conf}.  Confidence
is generally much lower for the simulated data (with a median of $0.0086$ to
$0.0140$ for the real-world data).  Finding higher confidence values in the
real-world data, which are expected to contain associations, indicates that the
confidence measure is able to suppress noise. However, the plots in
Figure~\ref{fig:conf} also show that confidence always increases with the item
in the right hand side of the rule ($l_j$) getting more frequent. This behavior
directly follows from the way confidence is calculated.  If the frequency of
the right hand side of the rule increases, confidence will increase even if the
items in the rule are not related (see itemset $Y$ in Equation~\ref{equ:conf}).
For the Grocery data set in Figure~\ref{fig:conf}(b) we see that
this effect dominates the confidence measure.  The fact that confidence clearly
favors some rules makes the measure problematic when it comes to selecting or
ranking rules.

\subsection{The interest measure lift}
\label{sec:lift}

Typically, rules mined using minimum support (and confidence) are
filtered or ordered using their lift value.  The measure lift
(also called interest \citep{arules:Brin1997}) is defined on rules of the form
$X \Rightarrow Y$ as
\begin{equation}
\mathrm{lift}(X \Rightarrow Y) = \frac{\mathrm{conf}(X \Rightarrow Y)}{\mathrm{supp}(Y)}.
\end{equation}

A lift value of $1$ indicates that the items are co-occurring in the
database as expected under independence.  Values greater than one
indicate that the items are associated.  For marketing applications it
is generally argued that $\mathrm{lift} > 1$ indicates complementary products
and $\mathrm{lift} < 1$ 
indicates substitutes~\cite{Betancourt1990,Hruschka1999}.

\begin{figure}[pt]
\begin{minipage}[b]{.50\linewidth}
\centering
\includegraphics[width=\linewidth]{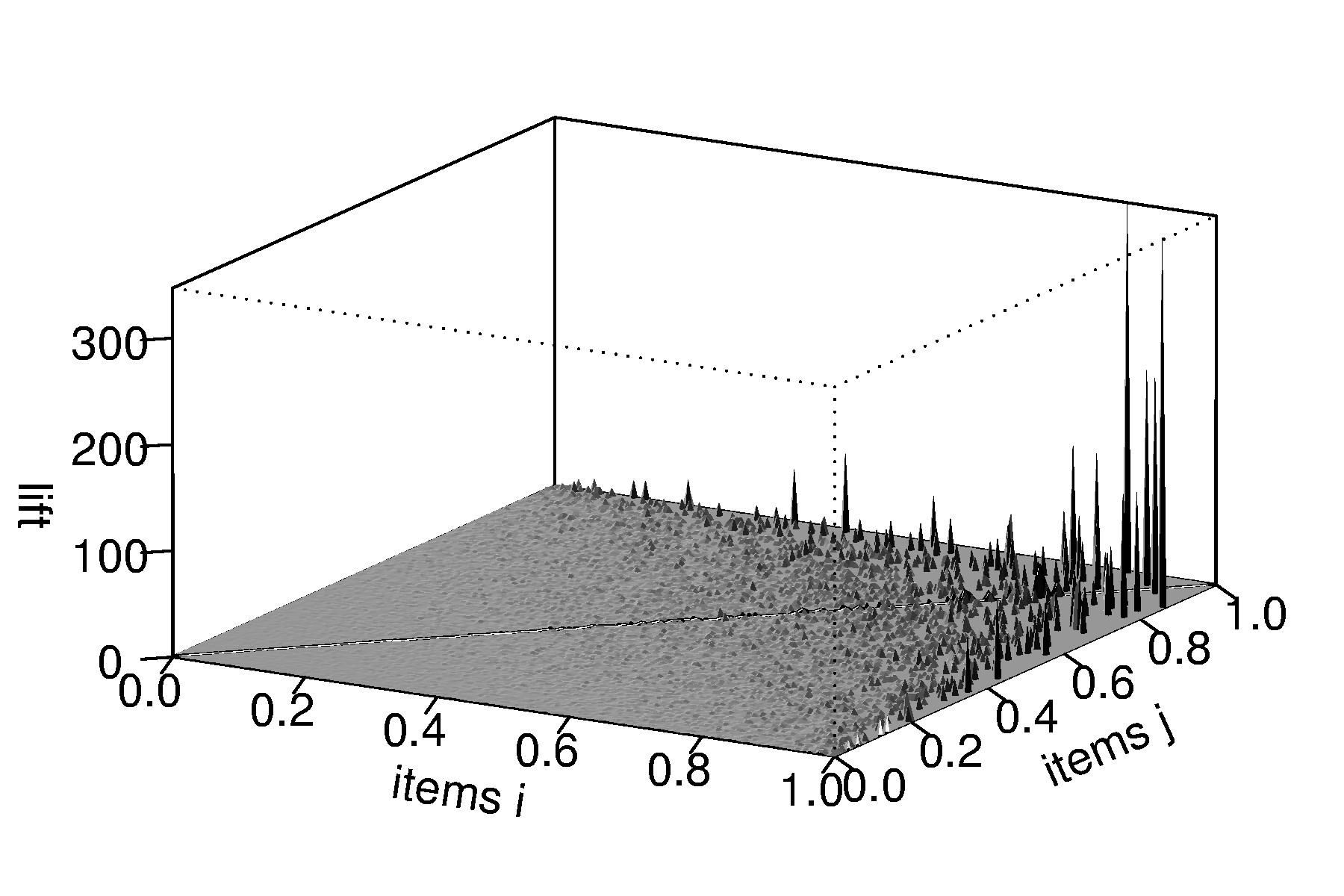}
(a) simulated
\end{minipage}
\begin{minipage}[b]{.50\linewidth}
\centering
\includegraphics[width=\linewidth]{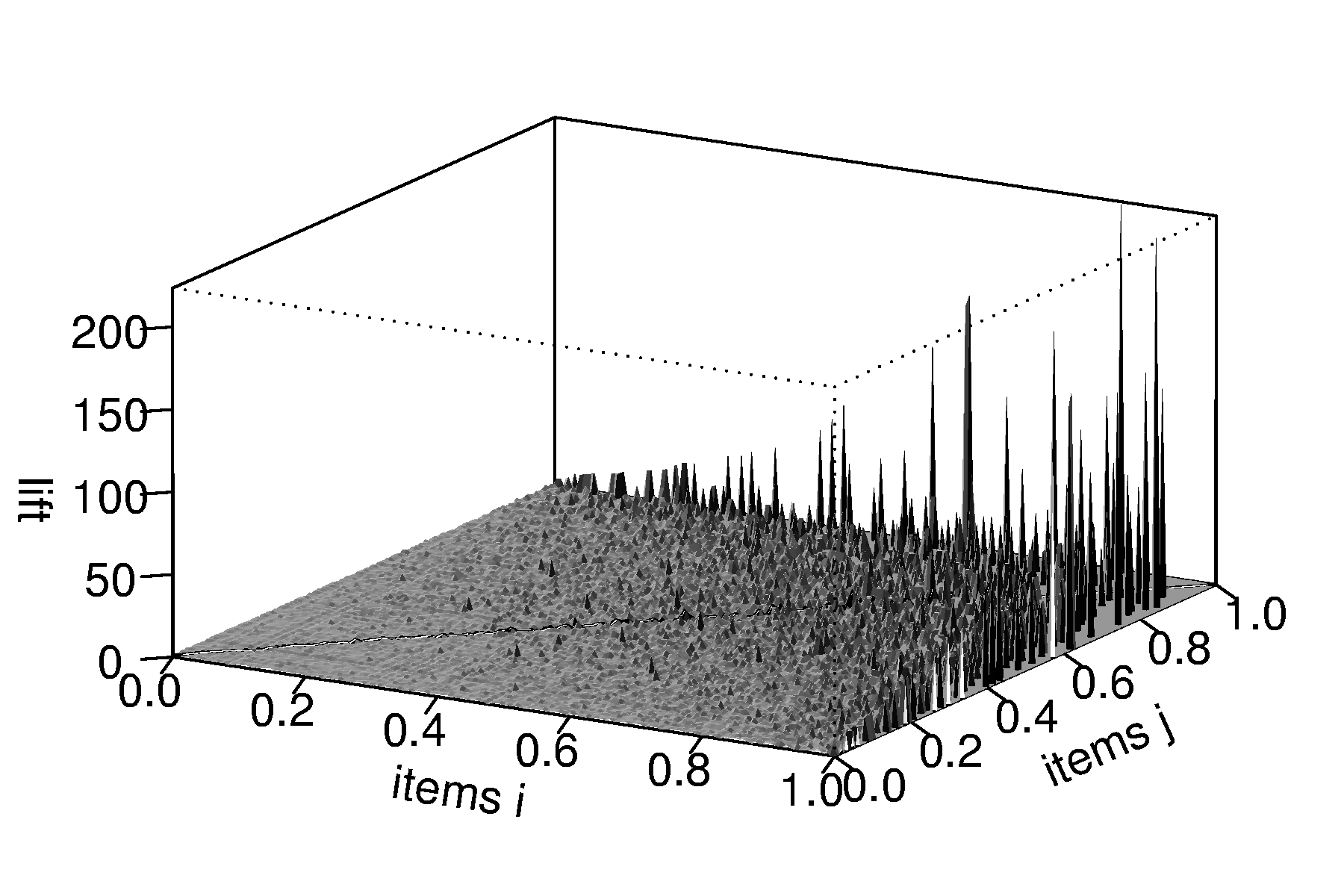}
(b) Grocery
\end{minipage}
\begin{minipage}[b]{.50\linewidth}
\centering
\includegraphics[width=\linewidth]{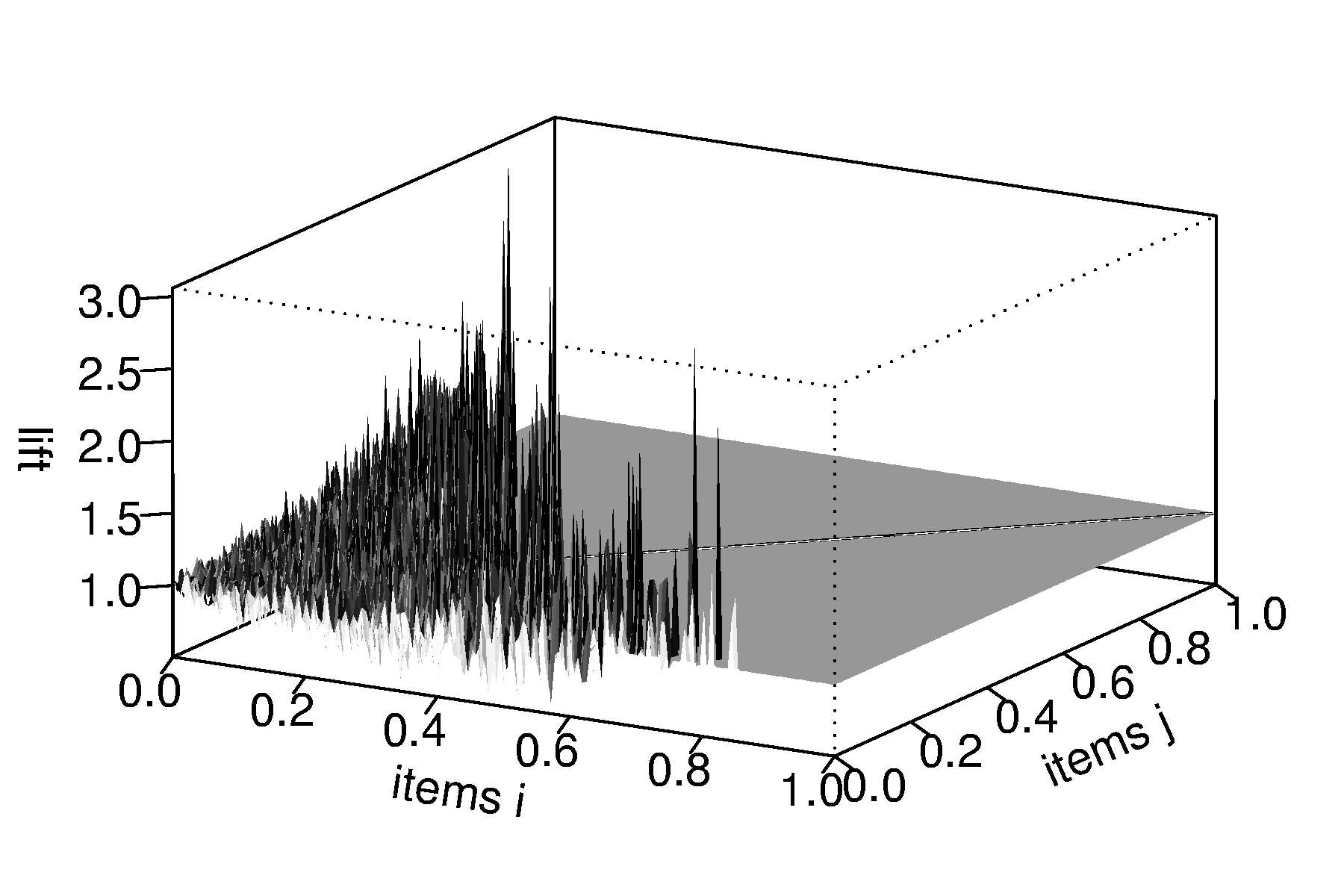}
(c) simulated with $\mathrm{supp}>0.1\%$
\end{minipage}
\begin{minipage}[b]{.50\linewidth}
\centering
\includegraphics[width=\linewidth]{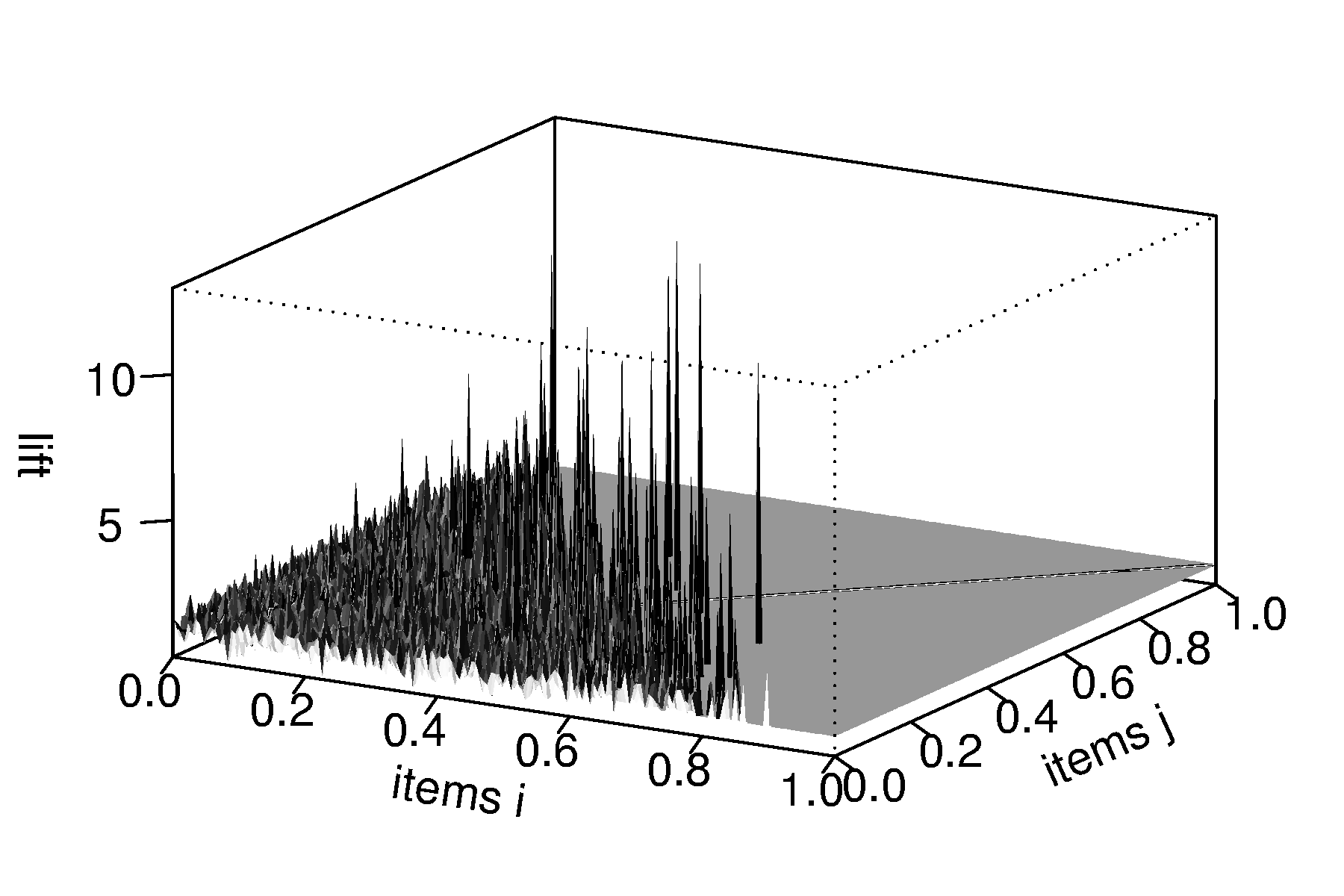}
(d) Grocery with $\mathrm{supp}>0.1\%$
\end{minipage}
\caption{Lift distributions of all rules with two items.\label{fig:lift}}
\end{figure}

Figure~\ref{fig:lift} 
show the lift values for the two data sets. The general distribution is
again very similar. In the plots in Figures~\ref{fig:lift}(a)
and \ref{fig:lift}(b) we can only see that very infrequent items 
produce extremely high lift values. These values are artifacts
occurring when two very rare items co-occur once together by chance.
Such artifacts are usually avoided in association rule mining by using a 
minimum support on itemsets. In Figures~\ref{fig:lift}(c) and
\ref{fig:lift}(d) we applied a minimum support of~0.1\%.
The plots show that there exist rules with higher lift values in the
Grocery data set 
than in the simulated data.
However, 
in the simulated data we still find
50~rules with a lift greater than~2.
 This indicates that the lift measure performs poorly to
 filter random noise in transaction data especially if we are also
 interested in relatively rare items with low support.
 The plots in Figures~\ref{fig:lift}(c) 
 and~\ref{fig:lift}(d)
 also clearly show lift's
 tendency to produce higher values for rules containing less
 frequent items resulting in that the highest lift values always
 occur close to the boundary of the selected minimum support.
 We refer the reader to \cite{arules:Bayardo1999} for a theoretical treatment
 of this effect.  If lift is used to rank discovered rules this means that
 there is not only a systematic tendency towards favoring rules with less
 frequent items but the rules with the highest lift will also always change 
 with even small variations of the user-specified minimum support.

\section{New measures of interest}
\label{sec:new_measures}

In the simple probabilistic model all items as well as combinations of items 
occur following independent Poisson processes.
If we look at the observed co-occurrence counts of all pairs of 
two items, $l_i$
and $l_j$, in a data set with $m$
transactions, we can form an $n\times n$ contingency table.
Each cell can be modeled by a random variable $C_{ij}$ which,
given fixed marginal counts $c_{i}$ and
$c_{j}$, follows a hyper-geometric distribution.

The hyper-geometric distribution
arises for the so-called urn problem, where the urn contains $w$
white balls and $b$ black balls.  
The number of white balls drawn with $k$
trials without replacement follows a hyper-geometric distribution.  This
model is applicable for counting co-occurrences for independent items
$l_i$ and $l_j$ in the following way: Item $l_j$
occurs in $c_{j}$ transactions, therefore, we can represent the database 
as an urn which contains $c_{j}$ transactions
with $l_j$ (white balls) and $m-c_{j}$ 
transactions without $l_j$ (black balls).
To assign item $l_i \ne l_j$ randomly to $c_{i}$ transactions,
we draw without replacement $c_{i}$
transactions from the urn. The number of drawn transactions
which we assign item $l_j$ to
(and thus represent the co-occurrences between $l_i$ and $l_j$)
then has a hyper-geometric distribution.

It is straightforward to extend this reasoning from two items to two 
itemsets $X$ and $Y$.
In this case the random variable $C_{XY}$ follows a hyper-geometric 
distribution with the counts of the itemsets as its parameter.
Formally, the probability of counting exactly $r$~transactions 
which contain the two independent itemsets $X$ and $Y$ is given by
\begin{equation}
  P(C_{XY} = r) = \frac{\binom{c_Y}{r} \binom{m-c_Y}{c_X-r}}{\binom{m}{c_X}}.
  \label{equ:hypergeometric}
\end{equation}
Note that this probability is conditional to the marginal counts
$c_X$ and $c_Y$. To simplify the notation, we 
will omit this condition also in the rest of the paper.

The probability of counting more than $r$ transactions is 
\begin{equation}
  P(C_{XY} > r) =  1 - \sum^r_{i=0} P(C_{XY} = i).
  \label{equ:hypergeometric_sum}
\end{equation}


Based on this probability, we will develop 
the probabilistic measures hyper-lift and hyper-confidence
in the rest of this section. Both measures quantify the deviation of the data 
from the independence model.
This idea is a similar to the use of random data 
to assess the significance of
found clusters in cluster analysis (see, e.g., \cite{misc:bock:1996}).

\subsection{Hyper-lift} 

The expected value of a random variable $C$ with a hyper-geometric
distribution is
\begin{equation}
  E(C) = \frac{k w}{w+b},
\end{equation}
where the parameter $k$ represents the number of trials, $w$ is the
number of white balls, and $b$ is the number of black balls.  Applied to
co-occurrence counts for the two itemsets $X$ and $Y$ in a transaction
database this gives
\begin{equation}
  E(C_{XY}) = \frac{c_X c_Y}{m},
\label{equ:expectedValue}
\end{equation}
where $m$ is the number of transactions in the database.  
By using Equation~\ref{equ:expectedValue} and the relationship between 
absolute counts and support, lift
can be rewritten as
\begin{equation}
  \mathrm{lift}(X \Rightarrow Y) = 
  \frac{\mathrm{conf}(X \Rightarrow Y)}{\mathrm{supp}(Y)} = 
  \frac{\mathrm{supp}(X \cup Y)}{\mathrm{supp}(X)\ \mathrm{supp}(Y)} =
  \frac{c_{XY}}{E(C_{XY})}.
\end{equation}

For items with a relatively high occurrence frequency, using the expected
value for lift works well.  However, for relatively infrequent items,
which are the majority in most transaction databases and very common
in other domains~\citep{arules:Xiong2003}, using the ratio of
the observed count to the expected value is problematic.  For example,
let us assume that we have the two independent itemsets $X$ and $Y$,
and both itemsets have a support of 1\% in a database with 10000~transactions. 
Using Equation~\ref{equ:expectedValue}, the expected count $E(C_{XY})$ is 1.
However, for the two independent itemsets 
there is a $P(C_{XY} > 1)$ of 0.264
(using the hyper-geometric distribution from 
Equation~\ref{equ:hypergeometric_sum}).
%
%
Therefore there is a substantial chance that we will see a lift value of
$2, 3$ or even higher.  Given the huge number of itemsets and rules
generated by combining items (especially when also considering itemsets
containing more than two items), this is very problematic. Using larger
databases with more transactions reduces the problem. 
However, it is not always possible to obtain a consistent database of
sufficient size. Large databases are usually collected over a long period of
time and thus may contain outdated information. For example, in a supermarket
the articles offered may have changed or shopping behavior may have changed due
to seasonal changes.

To address the problem, one can 
quantify the deviation of the observed co-occurrence count $c_{XY}$ from
the independence model by dividing it by a different location parameter
of the underlying hyper-geometric distribution than the mean which is used 
for lift.
For hyper-lift we suggest to use the quantile of the distribution denoted
by $Q_{\delta}(C_{XY})$.  
Formally, the minimal value of the $\delta$ quantile of the distribution 
of $C_{XY}$ is defined by the following inequalities:
\begin{equation}
  P(C_{XY} < Q_{\delta}(C_{XY})) \le \delta 
  \quad \text{and} \quad
P(C_{XY} > Q_{\delta}(C_{XY})) \le 1 - \delta.  
\label{equ:quantile}
\end{equation}
The resulting measure, which we call  hyper-lift, 
is defined as
\begin{equation}
\mathrm{hyper\mbox{-}lift_\delta}(X \Rightarrow Y) = \frac{c_{XY}}{Q_{\delta}(C_{XY})}.
\label{equ:hyper-lift}
\end{equation}

In the following, we will use $\delta = 0.99$ which results in hyper-lift
being more conservative compared to lift.  The measure can be
interpreted as the number of times the observed co-occurrence count
$c_{XY}$ is higher than the highest count we expect at most 99\% of the
time.  This means, that hyper-lift for a rule with independent items
will exceed 1 only in 1\% of the cases.

\begin{figure}[pt]
  \begin{minipage}[b]{.50\linewidth}
    \centering
    \includegraphics[width=\linewidth]{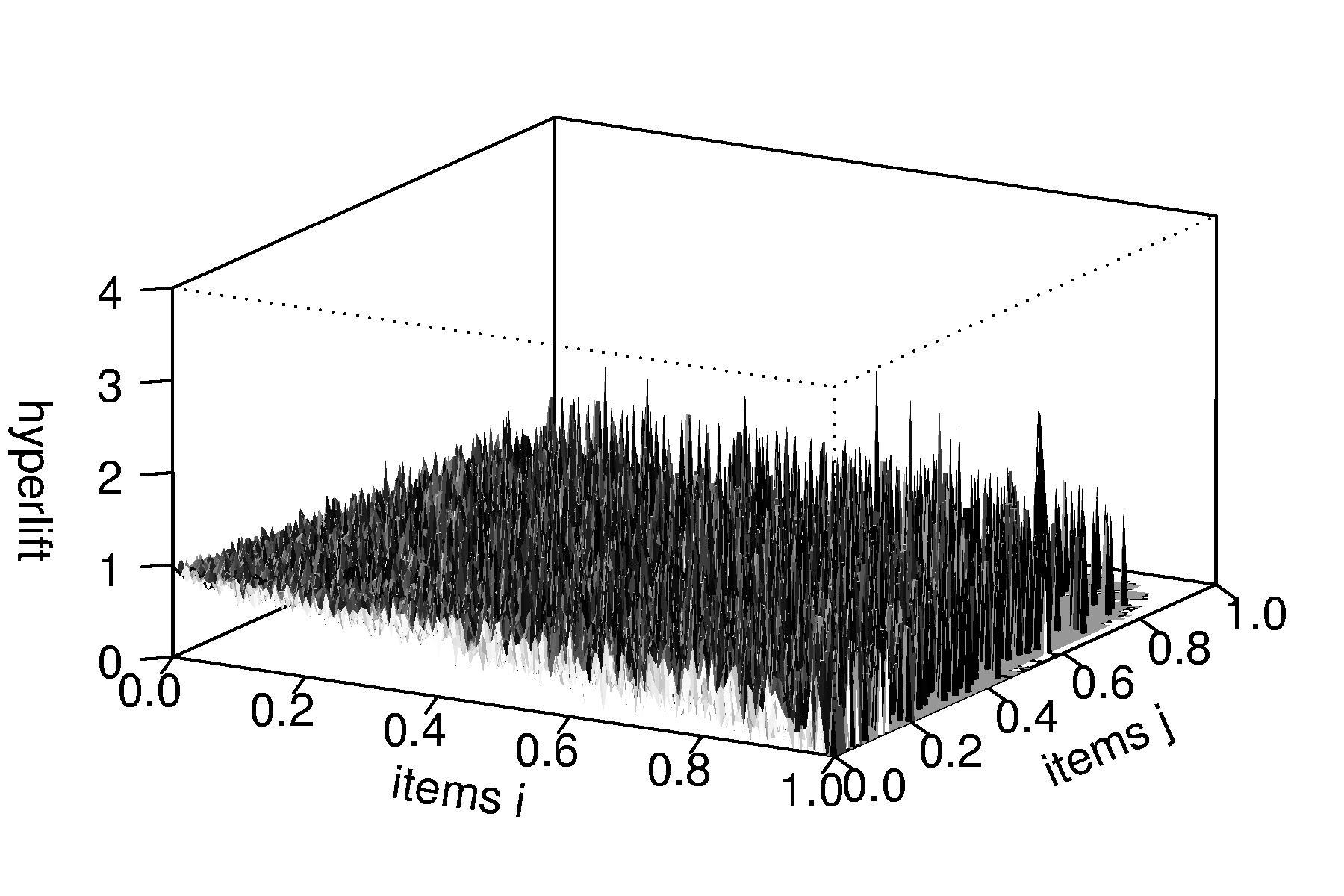}
    (a) simulated
\end{minipage}
\begin{minipage}[b]{.50\linewidth}
  \centering
  \includegraphics[width=\linewidth]{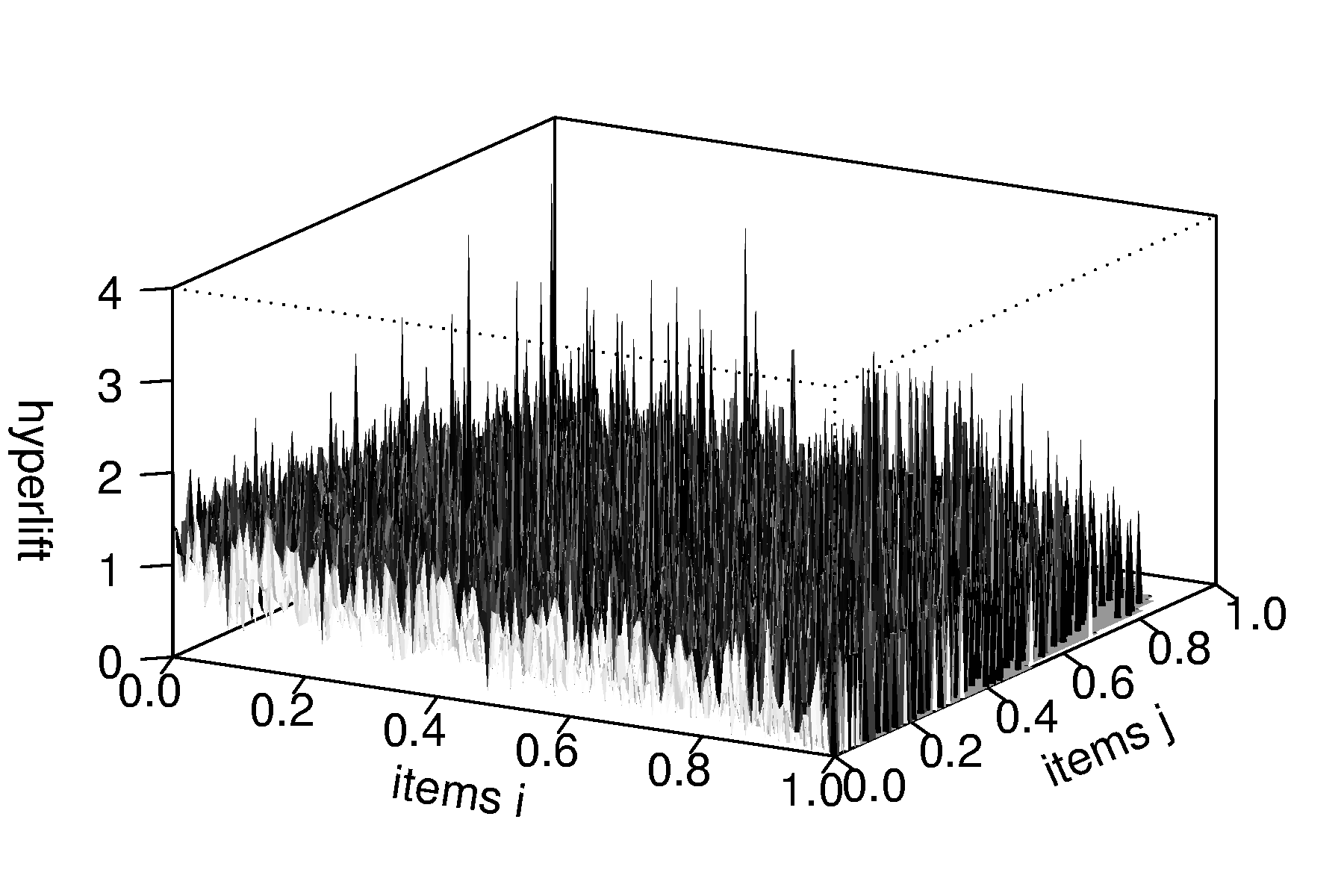}
  (b) Grocery
\end{minipage}
    \caption{Hyper-lift for rules with two items.\label{fig:hyperlift}}
\end{figure}

In Figure~\ref{fig:hyperlift} we compare the 
distribution of the
hyper-lift values for all rules with two items at $\delta = 0.99$
for the simulated and the Grocery database.
Figure~\ref{fig:hyperlift}(a) shows that the
hyper-lift on the simulated data is more evenly
distributed than lift (compare to Figure~\ref{fig:lift} in 
Section~\ref{sec:lift}). Also only for 
100 of the $n\times n= 28561$ rules hyper-lift exceeds 1 and no rule 
exceeds 2. This indicates that
hyper-lift filters the random co-occurrences better than lift with
3718 rules having a lift greater than 1 and 82 rules exceed a lift of 2.
However, hyper-lift also shows a systematic dependency
on the occurrence probability of items leading to smaller and more volatile
values for rules with less frequent items.

On the Grocery database in Figure~\ref{fig:hyperlift}(b) 
we find larger hyper-lift
values of up to 4.286. This indicates that the Grocery database 
indeed contains
dependencies. The highest values are observed between items with intermediate
support (located closer to the center of the plot). Therefore, hyper-lift 
avoids lift's problem of producing the highest values always only close to
the minimum support boundary (compare Section~\ref{sec:lift}).

Further evaluations of hyper-lift with rules including an arbitrary
number of items will be presented in Section~\ref{comparison}.

\subsection{Hyper-confidence}
\label{sec:hyper-confidence}

Instead of looking at quantiles of the hyper-geometric distribution 
to form a lift-like measure, we 
can also directly calculate the probability of realizing a count
smaller than the observed co-occurrence count~$c_{XY}$
given the marginal counts $c_X$ and $c_Y$.

\begin{equation}
  P(C_{XY} < c_{XY}) = \sum_{i=0}^{c_{XY}-1} P(C_{XY} = i),
\end{equation}
where $P(C_{XY} = i)$ is calculated 
using Equation~\ref{equ:hypergeometric} 
above.
A high probability indicates
that observing $c_{XY}$ under independence is rather unlikely.
The probability can be directly used as the interest measure
hyper-confidence:
\begin{equation}
  \mathrm{hyper\mbox{-}confidence}(X \Rightarrow Y) =  P(C_{XY} < c_{XY})
  \label{equ:hyper_conf}
\end{equation}

Analogously to other measures of interest, we can use a 
threshold $\gamma$ on hyper-confidence to accept 
only rules for which the probability to observe such a high co-occurrence count
by chance
is smaller or equal than $1- \gamma$. For example, if we use
$\gamma = 0.99$, for each accepted rule, there is only a 1\% chance
that the observed co-occurrence count arose by pure chance.
Formally, using a threshold on hyper-confidence 
for the rules $X \Rightarrow Y$ (or $Y \Rightarrow X$) can be 
interpreted as using a one-sided statistical test on the $2 \times 2$
contingency table depicted in Table~\ref{tab:2x2_contingency_table}
with the null hypothesis that $X$ and $Y$ are not
positively related.
It can be shown that 
hyper-confidence is related to the
$p$-value of a one-sided Fisher's exact 
test.
The one-sided Fisher's exact test for $2 \times 2$ contingency tables 
is a simple permutation test which evaluates the probability
for realizing any table 
(see Table~\ref{tab:2x2_contingency_table}) 
with $C_{XY} \ge c_{XY}$ given fixed marginal counts~\citep{misc:Fisher:1935}.
The test's $p$-value is given by
\begin{equation}
  p\mathrm{\mbox{-}value} = P(C_{XY} \ge c_{XY}) 
\end{equation}
which is equal to $1 - \mathrm{hyper\mbox{-}confidence}(X \Rightarrow
Y)$ (see Equation~\ref{equ:hyper_conf}), and gives the $p$-value of the
uniformly most powerful (UMP) test for the null $\rho \le 1$ (where
$\rho$ is the odds ratio) against the alternative of positive
association $\rho > 1$ \citep[][pp.~58--59]{misc:lehmann:1959}, provided
that the $p$-value of a randomized test is defined as the lowest
significance level of the test that would lead to a (complete)
rejection.

If we use a significance level of $\alpha = 0.01$, we would reject the
null hypothesis of no positive correlation 
if $p\mathrm{\mbox{-}value} < \alpha$.
Using $\gamma$ as a threshold on hyper-confidence is equivalent to
a Fisher's exact test with $\alpha = 1 - \gamma$.

\begin{table}[tp]
\centering
\begin{tabular}{c|cc|c}
  & $X=0$			& $X=1$		    & \\
\hline
$Y=0$ & $m-c_Y-c_X - C_{XY}$	& $c_X - C_{XY}$    & $m-c_Y$ \\
$Y=1$ & $c_Y- C_{XY}$		&  $C_{XY}$	    & $c_Y$ \\
\hline
     & $m-c_X$			&  $c_X$		    & $m$ \\
\end{tabular}
\caption{$2 \times 2$ contingency table for the counts of the 
presence (1) and absence (0) of the itemsets in transactions.}
\label{tab:2x2_contingency_table}
\end{table}

Note that hyper-confidence is 
equivalent to a special case of Fisher's exact test, the one-sided 
test on $2 \times 2$ contingency tables. In this case, the $p$-value is 
directly obtained from the hyper-geometric distribution which is 
computationally negligible compared to the effort of counting support 
and finding frequent itemsets. 

The idea of using a statistical test on $2 \times 2$ contingency tables to test
for dependencies between itemsets was already proposed by
\citet{arules:Liu1999}. The authors use the $\chi^2$ test  which is an
approximate test for the same purpose as Fisher's exact test in the
2-sided case. The generally
accepted rule of thumb is that the $\chi^2$ test's approximation 
breaks down if the expected counts for any of the contingency table's cells
falls below $5$. For data mining applications, where potentially millions of
tests have to be performed, it is very likely that many tests will suffer from
this restriction.  Fisher's exact test and thus hyper-confidence do not have
this drawback. Furthermore, the $\chi^2$ test is a two-sided test, but for the 
application of mining association rules where only 
rules with positively correlated elements are of interest, a one-sided test
as used here is much more appropriate.

\begin{figure}[pt]
\begin{minipage}[b]{.50\linewidth}
  \centering
  \includegraphics[width=\linewidth]{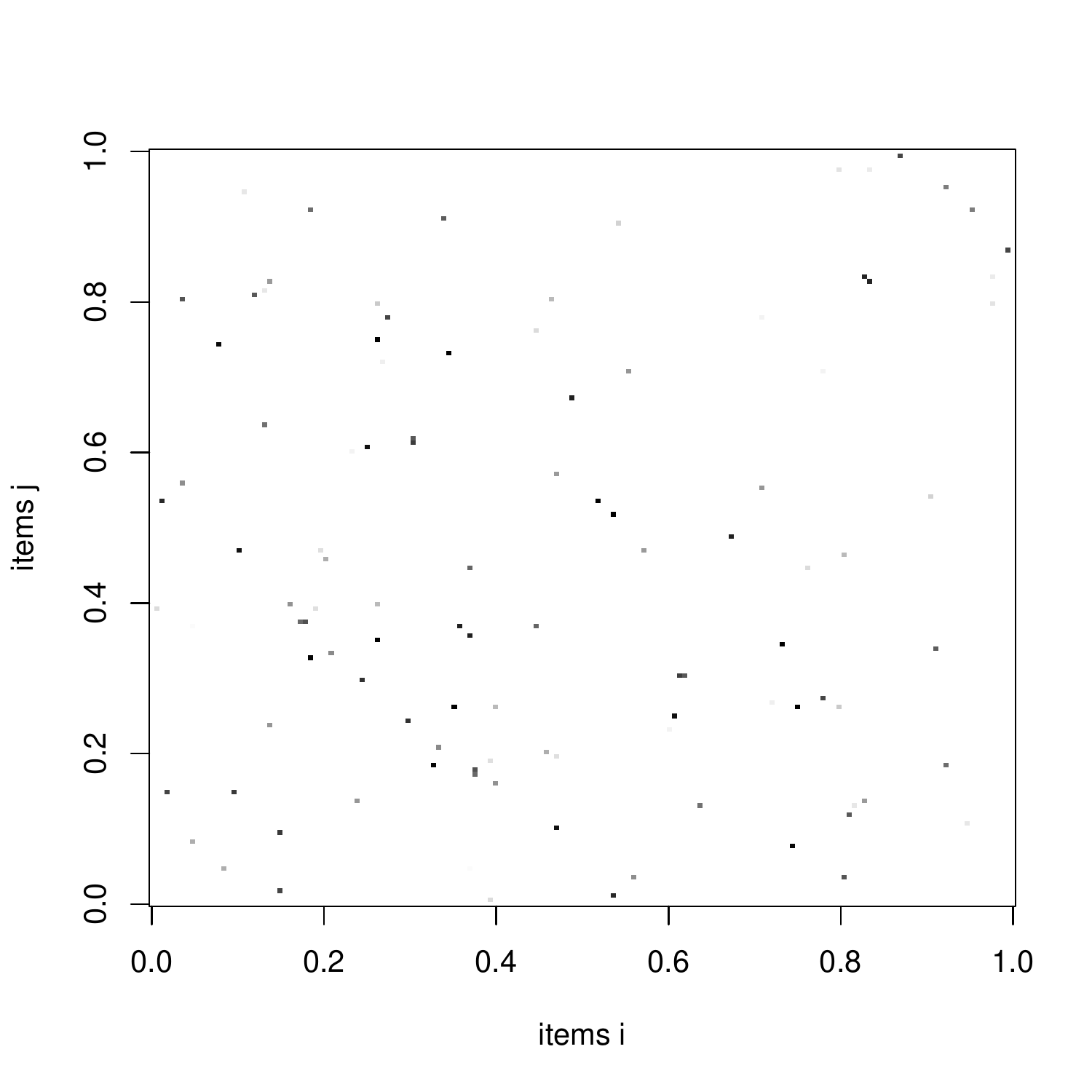}
  (a)  simulated 
\end{minipage}
  \begin{minipage}[b]{.50\linewidth}
    \centering
    \includegraphics[width=\linewidth]{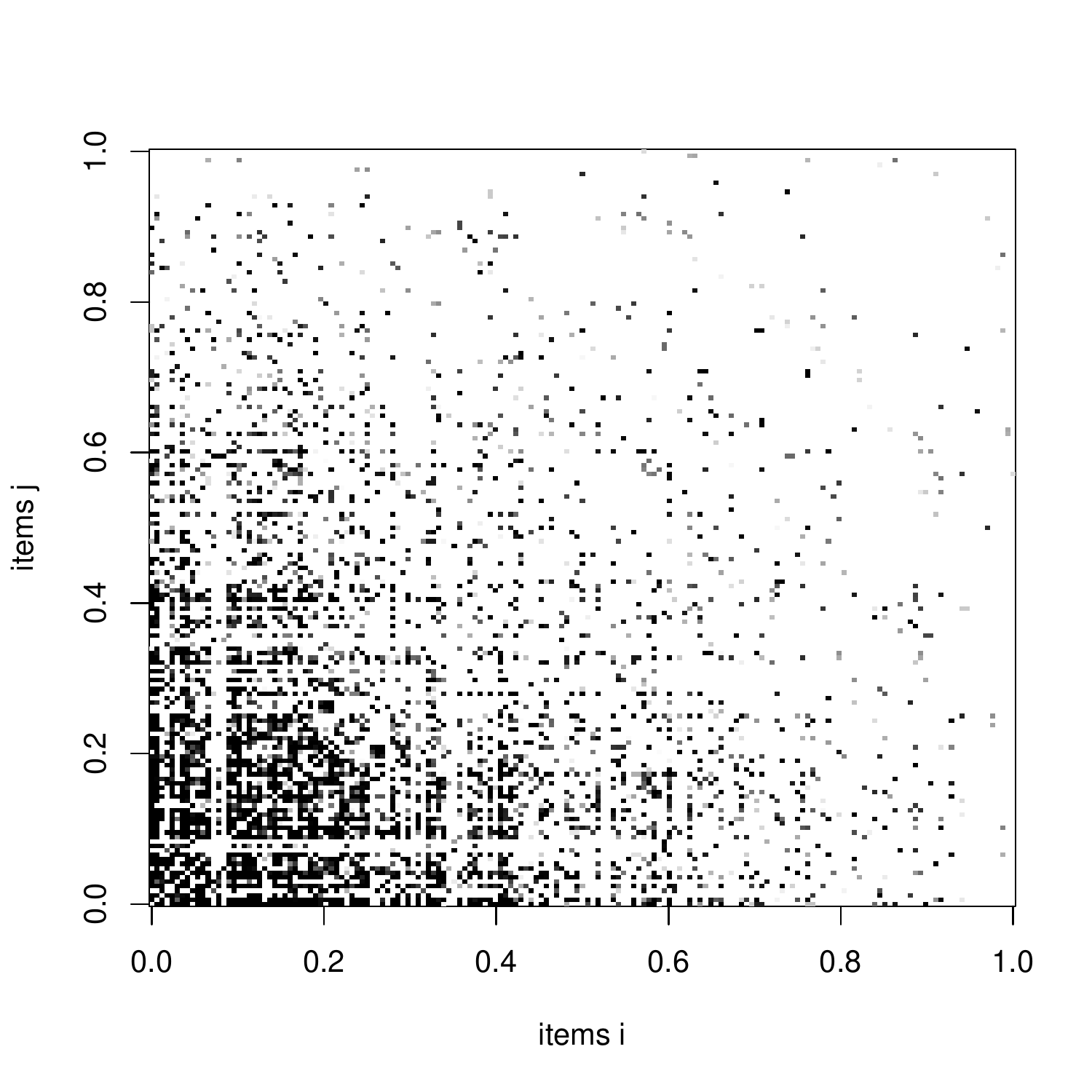}
    (b)   Grocery
  \end{minipage}
  \caption{Hyper-confidence for rules with two items 
  and $\gamma > 0.99$
  (items are ordered by decreasing support from left to right and 
  bottom to top).\label{fig:images}}
\end{figure}

In Figures~\ref{fig:images}(a) and~(b) we compare the 
hyper-confidence values 
produced for all rules with 2 items on the Grocery database and the 
corresponding simulated data set. 
Since the values vary strongly between 0 and 1, we use 
for easier comparison image plots instead of the perspective plots used 
before.
The intensity of the dots indicates the value of hyper-confidence
for the rules $l_i \Rightarrow l_j$ 
(the items are again organized left to right and front to back by 
decreasing support).
All dots for rules with a hyper-confidence value smaller than a set 
threshold of $\gamma = 0.99$ are removed.
For the simulated data we see that the 108 rules which pass the 
hyper-confidence threshold are scattered over the whole image.
For the Grocery database in Figure~\ref{fig:images}(b) we see that many 
(3732) rules pass the hyper-confidence threshold and that the concentration of
passing rules increases with item support. This results from the fact
that with increasing counts the test is better able to reject the 
null hypotheses.

\begin{figure}[pt]
\begin{minipage}[b]{.50\linewidth}
    \centering
    \includegraphics[width=\linewidth]{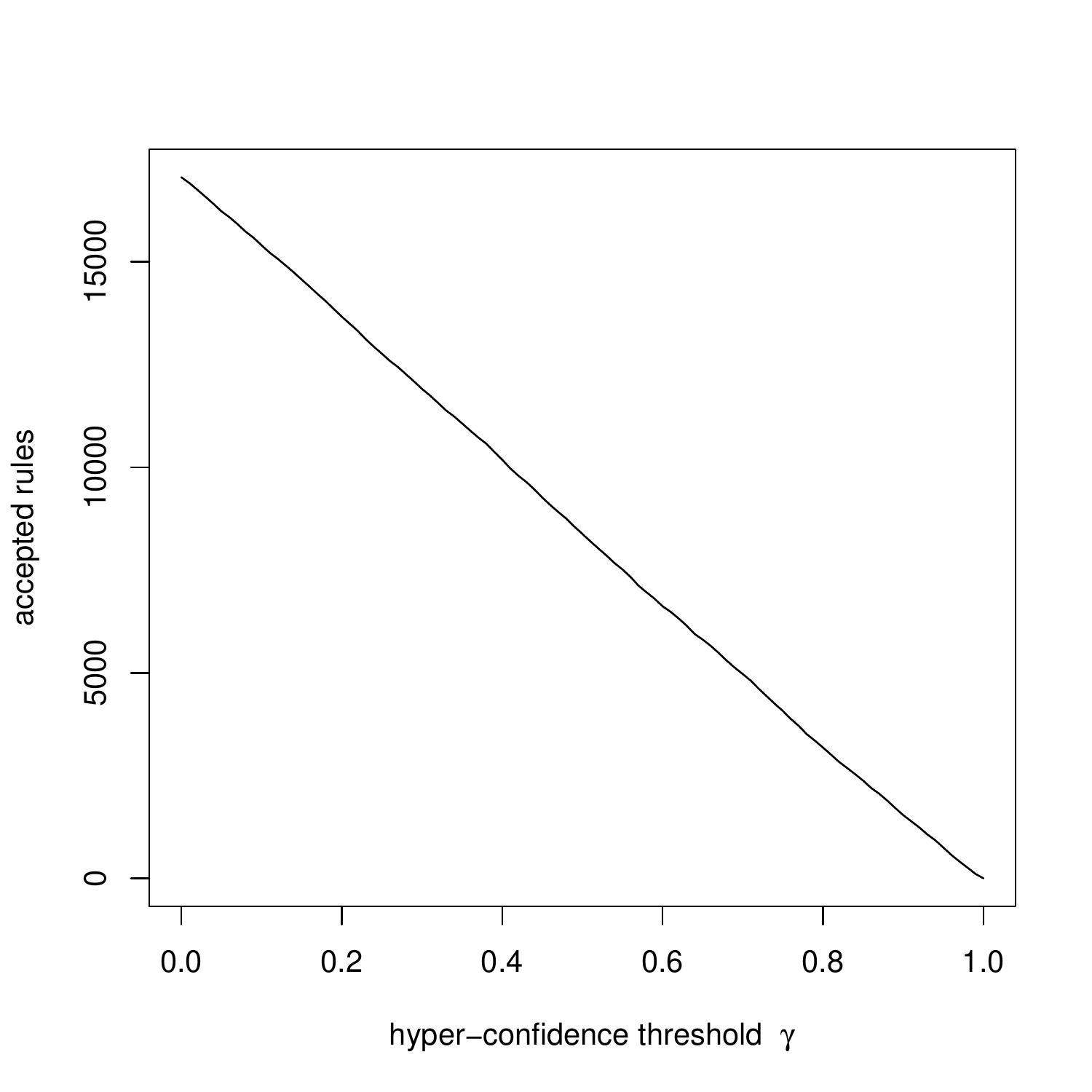}
    (a)  simulated
\end{minipage}
\begin{minipage}[b]{.50\linewidth}
    \centering
    \includegraphics[width=\linewidth]{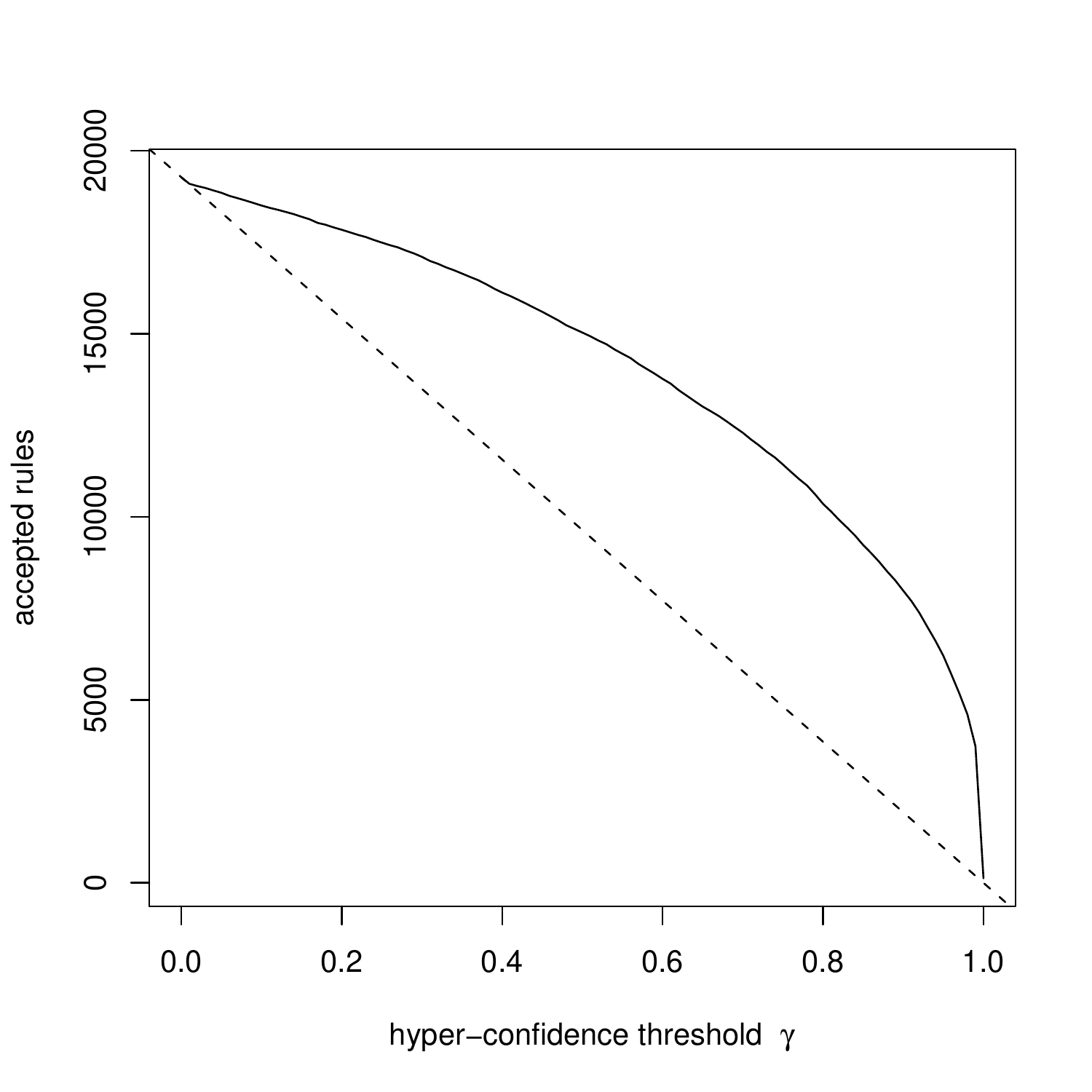}
    (b)  Grocery
\end{minipage}
  \caption{Number of accepted rules depending on the used 
  hyper-confidence threshold.\label{fig:hyper_conf}}
\end{figure}

\begin{figure}[pt]
\begin{minipage}[b]{.50\linewidth}
    \centering
    \includegraphics[width=\linewidth]{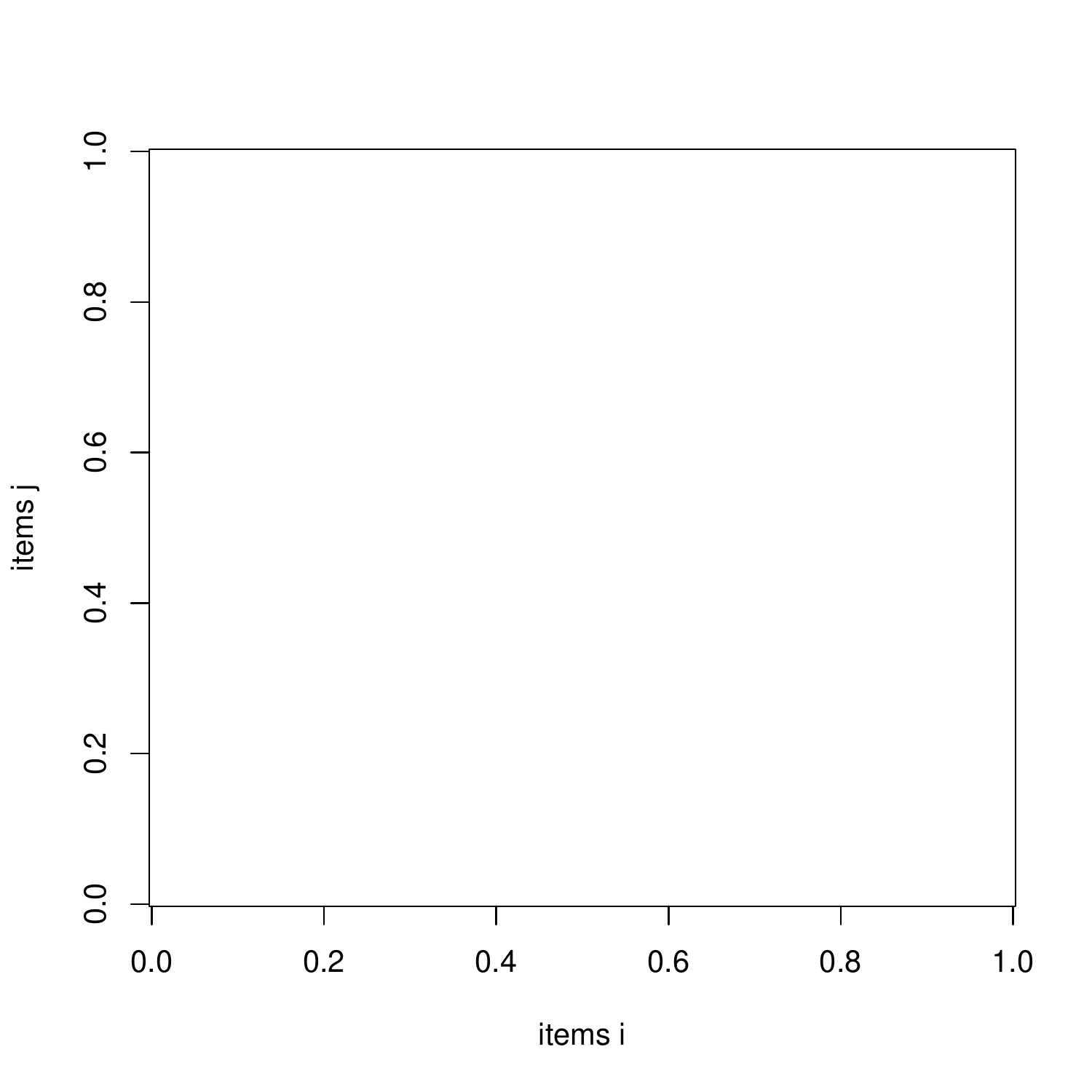}
    (c)  simulated
\end{minipage}
\begin{minipage}[b]{.50\linewidth}
    \centering
    \includegraphics[width=\linewidth]{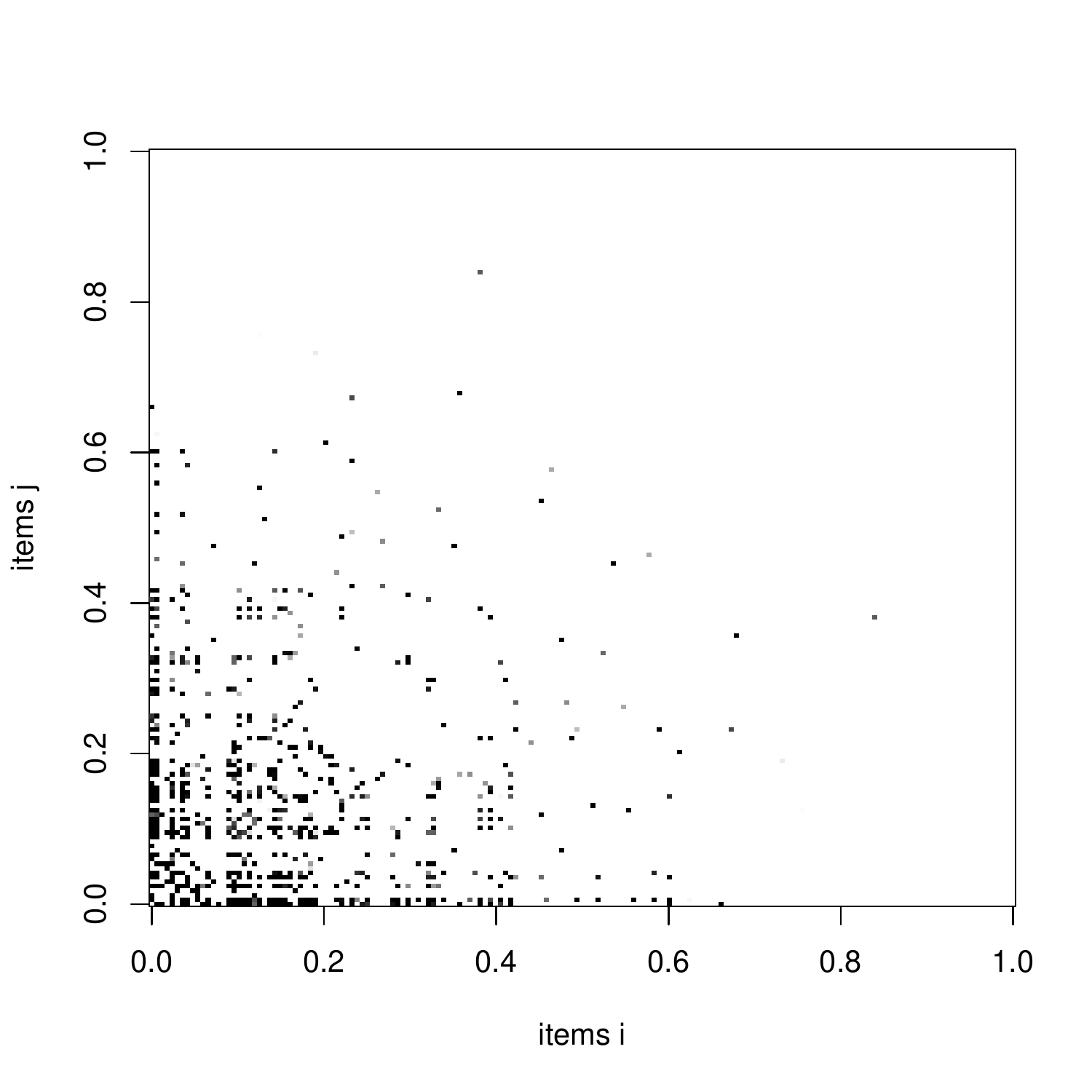}
    (d)  Grocery
\end{minipage}
  \caption{Hyper-confidence for rules with two items
  using a Bonferroni corrected~$\gamma$.\label{fig:images_bonf}}
\end{figure}

In Figures~\ref{fig:hyper_conf}(a) and~(b)
we present the number of accepted rules by the set hyper-confidence threshold.
For the simulated data the number of accepted rules is directly proportional
to $1- \gamma$. This behavior directly follows from the properties
of the data.
All items are independent and therefore
rules randomly surpass the threshold with the probability given by the 
threshold.
For the Grocery data set in Figure~\ref{fig:hyper_conf}(b),
we see that more rules than expected for random data (dashed line) 
surpass the threshold. At $\gamma = 0.99$, for each of the $n$ tests
exists a 1\% chance that the rule is accepted although it is spurious.
Therefore, a rough estimate of the proportion of 
spurious rules in the set of $m$ accepted
rules is $n(1-\gamma)/m$. 
For example, for the Grocery database we have $n = 19272$ tests and
for $\gamma = 0.99$ we found $m = 3732$ rules. The estimated proportion
of spurious rules in the set is therefore 5.2\% which is about five times
higher than the $\alpha$ of 1\% used for each individual test.
The reason is that we conduct multiple tests 
simultaneously to generate the set of accepted 
rules. If we are not interested in the individual test but in 
the probability that some tests
will accept spurious rules, we have to adjust $\alpha$. A
conservative approach is the Bonferroni correction~\citep{Shaffer1995} 
where a corrected
significance level of 
$\alpha^* = \alpha / n$ is used for each test to achieve an overall 
alpha value of $\alpha$. The result of using 
a Bonferroni corrected $\gamma = 1-\alpha^*$ is
depicted in Figures~\ref{fig:images_bonf}(a) and~(b).
For the simulated data set we see that after correction no spurious rule is
accepted while for the Grocery database still 652 rules pass the test.
Since we used a corrected threshold of $\gamma = 0.99999948$ 
($\alpha = 5.2 \cdot 10^{-7}$) these rules
represent very strong associations.

Between the measures hyper-confidence and hyper-lift 
exists a direct connection.
Using a threshold $\gamma$ on hyper-confidence and requiring a hyper-lift
using the $\delta$ quantile to be greater than one is equivalent for
$\gamma = \delta$. Formally, for an arbitrary rule $X \Rightarrow Y$ and
$0 < \gamma < 1$, 
\begin{equation}
\mathrm{hyper\mbox{-}confidence}(X \Rightarrow Y) \ge \gamma
\Leftrightarrow 
\mathrm{hyper\mbox{-}lift}_{\delta}(X \Rightarrow Y) > 1 
\quad \text{for} \quad 
\gamma = \delta.
\end{equation}

To prove this equivalence, let us write $F(c) = P(C_{XY} \le c)$ for the
distribution function of $C_{XY}$, so that the hyper-confidence~$h =
h(c)$ equals $F(c-1)$.  We note that $F$ as well as its quantile
function~$Q$ are non-decreasing and that for integer~$c$ in the support
of $C_{XY}$, $Q_{F(c)} = c$.  Hence, provided that $h(c) > 0$, $Q_{h(c)}
= c - 1$.  What we need to show is that $h(c) \ge \gamma$ iff $c >
Q_\gamma$.  If $0 < \gamma \le h(c)$, it follows that $Q_\gamma \le
Q_{h(c)} = c - 1$, i.e., $c > Q_\gamma$.  Conversely, if $c > Q_\gamma$,
it follows that $c - 1 \ge Q_\gamma$ and thus $h(c) = F(c-1) \ge
F(Q_\gamma) \ge \gamma$, completing the proof.

%
%
%
%

Hyper-confidence as defined above only uncovers complementary effects
between items.  To interpret using a threshold on hyper-confidence as a
simple one-sided statistical test makes it also very natural to adapt
the measure to find substitution effects, items which co-occur
significantly less together than expected under independence.  The
hyper-confidence for substitutes is given by:
\begin{equation}
  \mathrm{hyper\mbox{-}confidence}^\mathrm{sub}(X \Rightarrow Y) =  
  P(C_{XY} > c_{XY}) = 1- \sum_{i=0}^{c_{XY}} P(C_{XY} = i)
\end{equation}

Applying a threshold $\gamma^\mathrm{sub}$ can be again interpreted as
using a one-sided test, this time for negatively related items.  One
could also construct a hyper-lift measure for substitutes using low
quantiles; however, its construction is less straightforward.

\begin{figure}[pt]
\begin{minipage}[b]{.50\linewidth}
  \centering
  \includegraphics[width=\linewidth]{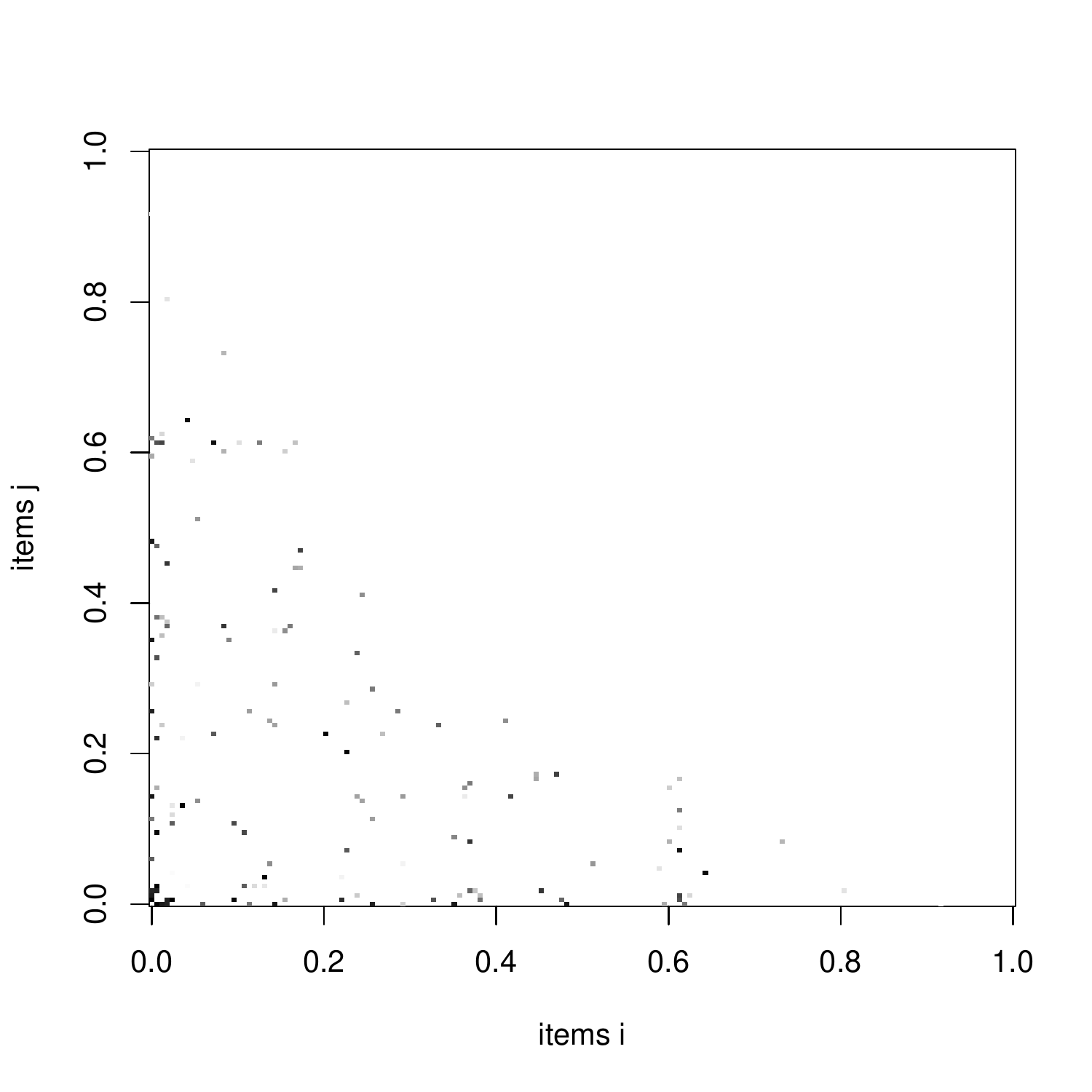}
  \centerline{(a) simulated}
\end{minipage}
\begin{minipage}[b]{.50\linewidth}
    \centering
    \includegraphics[width=\linewidth]{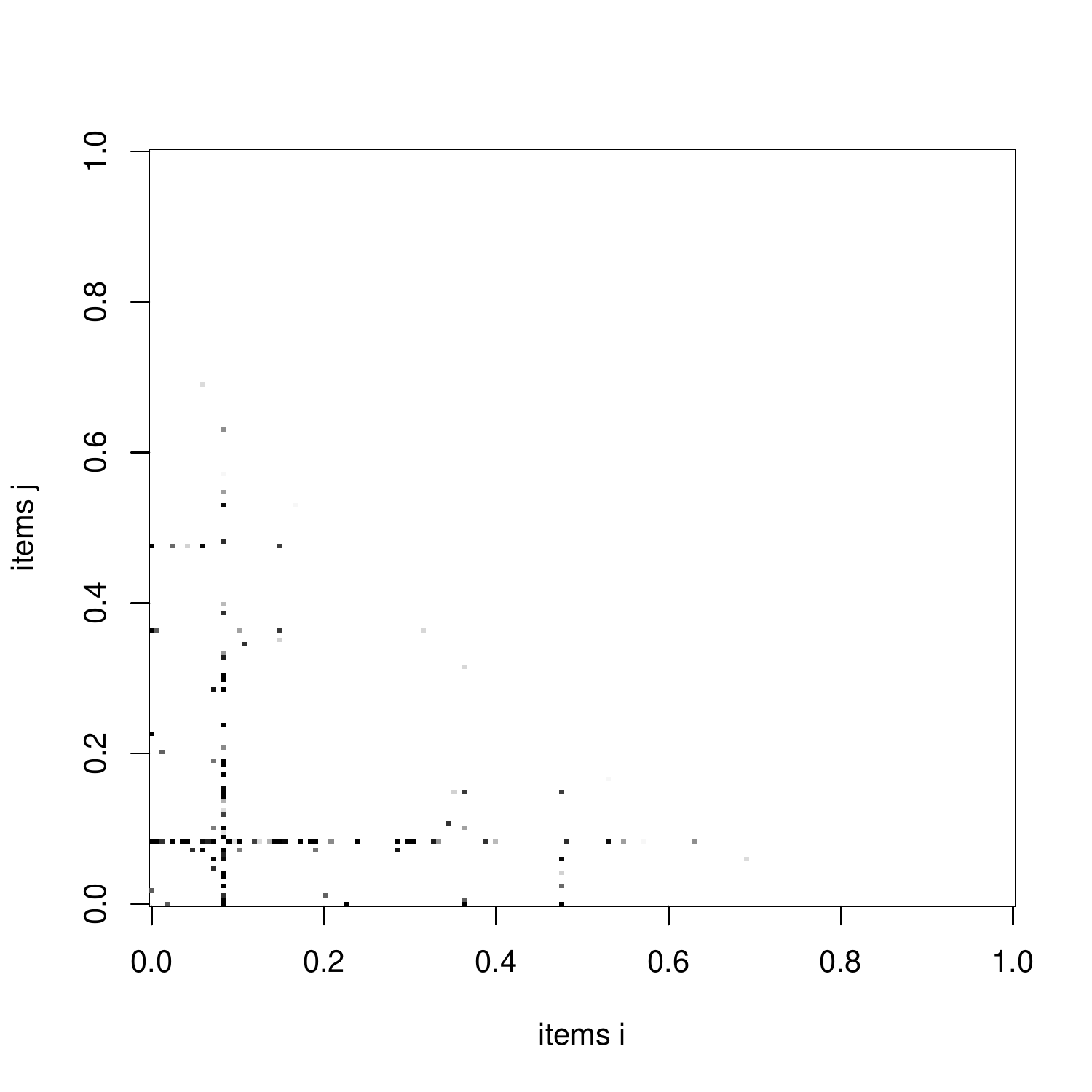}
  \centerline{(b) Grocery}
\end{minipage}
  \caption{Hyper-confidence for substitutes with $\gamma^\mathrm{sub} > 0.99$ 
  for rules with two items.\label{fig:image_subs}}
\end{figure}

In Figures~\ref{fig:image_subs}(a) and~(b) we show the rules with 
two items which 
surpass a $\gamma^\mathrm{sub}$ of 0.99 in the simulated and the
Grocery database. In the simulated data we see that the 68 falsely found
rules are regularly scattered over the lower left triangle.
In the Grocery database, the 116~rules which contain substitutes
are concentrated for a few items 
(the line clearly visible in Figure~\ref{fig:image_subs}(b) corresponds to
the item `canned beer' which has a strong substitution effect for most other
items). As for complements, it is possible to use Bonferroni correction.

\subsection{Empirical results}
\label{comparison}

To evaluate the proposed measures, we compare their ability 
to suppress spurious rules with the well-known lift measure.
For the evaluation, we use in addition to the Grocery database  
two publicly available databases. 
The database ``T10I4D100K'' is an artificial
data set generated by the procedure described by \citet{arules:Agrawal1994}
which is often used to evaluate association rule mining algorithms.
The third database is a sample of 50,000 transactions from the ``Kosarak''
database. This database was provided by \citet{misc:bodon:2003} and contains 
click-stream data of a Hungarian on-line news portal.
As shown in Table~\ref{tab:datasets}, 
the three databases 
have very different characteristics and thus
should cover a wide range of applications
for data from different sources and with different database sizes
and numbers of items.

\begin{table}[tp]
  \centering
  \begin{tabular}{lccc}
\hline
Database & Grocery & T10I4D100K & Kosarak \\
\hline
Type & market basket & artificial & click-stream \\
Transactions & 9835 & 100,000  & 50,000 \\ 
Avg. trans. size & 4.41 & 10.10 & 3.00\\
Median trans. size & 3.00 & 10.00 & 7.98\\
Distinct items & 169 & 870  & 18,936\\
\hline
\end{tabular}
  \caption{Characteristics of the used databases.\label{tab:datasets}}
\end{table}

For each database we simulate a comparable association-free data set 
following the simple probabilistic model described above in this paper.
We generate all rules with one item in the right hand side 
which satisfy a specified minimum support 
(see Table~\ref{tab:comp_lift_hyperlift}). 
Then we compare 
the impact of lift and confidence with  hyper-lift 
and hyper-confidence on rule selection.
In Table~\ref{tab:comp_lift_hyperlift} we present the 
number of rules found using the preset minimum support and the 
number of rules which also have 
a lift greater than $1$ and $2$, a hyper-lift with $\delta = 0.99$ greater 
than $1$ and $2$, or a hyper-confidence greater than $0.9$ and $0.999$,
respectively.
From the results in the table we see that,
compared to the real databases, 
in the simulated data sets only a much smaller number of rules 
reaches the required minimum support. This supports the assumption that
these data sets do not contain associations between items while 
the real databases do.
If we assume that rules found in the real databases are (at least potentially)
useful associations while we know that rules found in the simulated data sets
must be spurious, we can compare the performance of lift, 
hyper-lift and hyper-confidence on the data. 
In Table~\ref{tab:comp_lift_hyperlift} we see that 
there obviously exists a trade-off between accepting more rules in
the real databases while suppressing the spurious rules in the
simulated data sets. In terms of rules found in the real databases versus
rules suppressed in the simulated data sets, 
$\mathrm{hyper\mbox{-}lift_{0.99}} > 1$ lies 
for all three databases between $\mathrm{lift} > 1 $ and $\mathrm{lift} > 2 $
while $\mathrm{hyper\mbox{-}lift_{0.99}} > 2$ never accepts spurious rules
but also reduces the rules in the real databases (especially in the
Grocery database). The same is true for hyper-confidence with a
threshold of $0.9$ the number of resulting rules lying in between the 
results for the two lift thresholds and for $0.999$ hyper-confidence 
only once (for the Grocery database) accepts a single rule.

\begin{table}[tp]
  \centering
  \begin{tabular}{lr@{/}rr@{/}rr@{/}r}
\hline
Database  & Grocery&sim.  & T10I4D100K&sim. & Kosarak&sim.\\
\hline
Min. support & \multicolumn{2}{c}{0.001} & \multicolumn{2}{c}{0.001}& \multicolumn{2}{c}{0.002}   \\
Found rules			& 40943&8685 & 89605&9288   & 53245&2530\\
$\mathrm{lift > 1}$	& 40011&5812 & 86855&5592   & 51822&1365\\
$\mathrm{lift > 2}$	& 27334&180  & 84880&0	    & 42641&0\\
$\mathrm{hyper\mbox{-}lift}_{0.99} > 1$ & 30083&196  & 86463&150    & 51151&23\\
$\mathrm{hyper\mbox{-}lift}_{0.99} > 2$ & 1563&0  & 83176&0    &  37683&0 \\
$\mathrm{hyper\mbox{-}conf.} > 0.9$ & 36724&1531  & 86647&1286    & 51282&240\\
$\mathrm{hyper\mbox{-}conf.} > 0.9999$ & 15046&1  & 86207&0    & 51083&0\\
\hline
\end{tabular}
  \caption{Number of rules exceeding a 
  lift and hyper-lift ($\delta = 0.99$) of 1 and 2,
  and a hyper-confidence of 0.9 and 0.9999
  on the three databases and  
  comparable simulated 
  data sets.\label{tab:comp_lift_hyperlift}}
\end{table}


\begin{figure}[pt]
\begin{minipage}[b]{.50\linewidth}
  \centering
  \includegraphics[width=\linewidth]{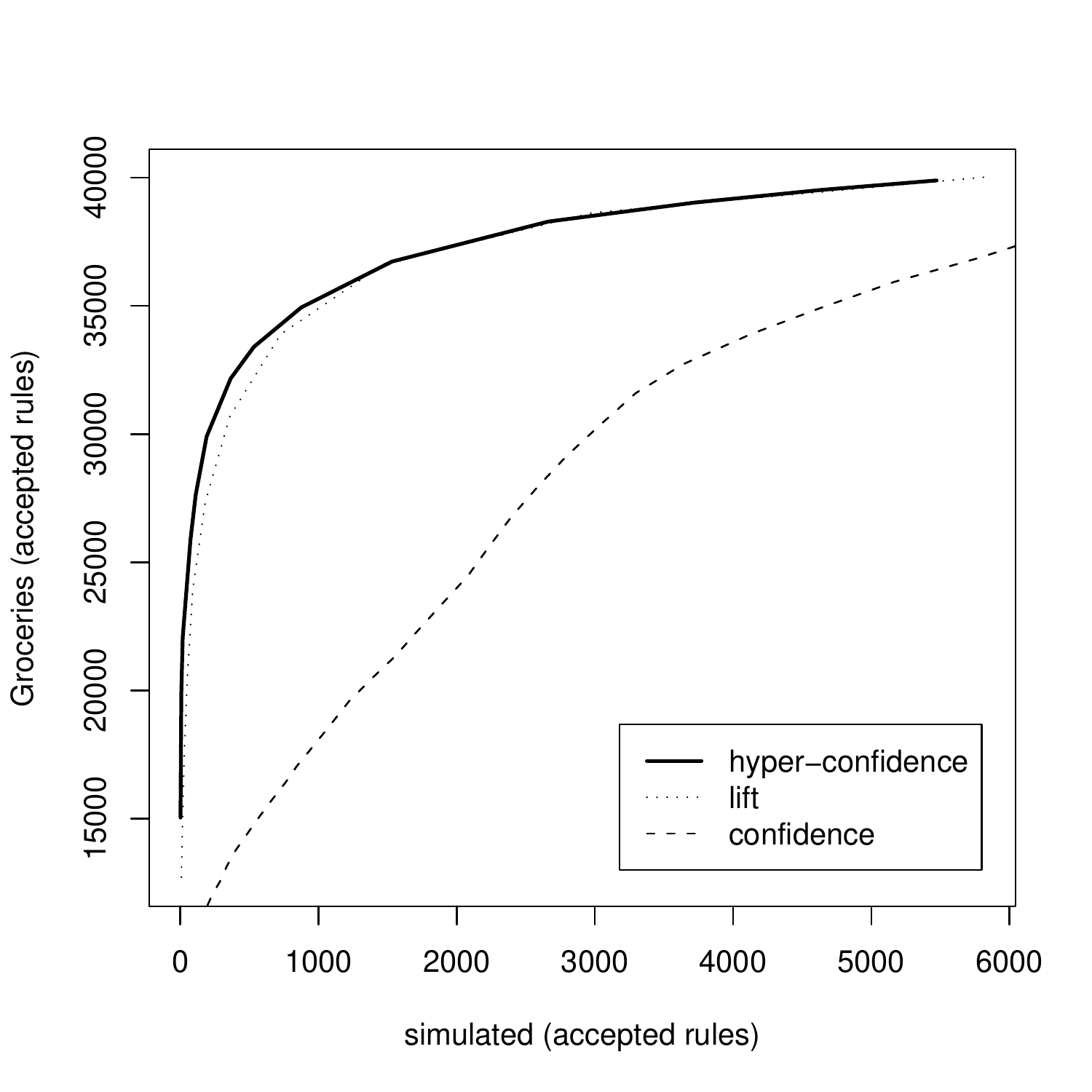}
\end{minipage}
  \begin{minipage}[b]{.50\linewidth}
    \centering
    \includegraphics[width=\linewidth]{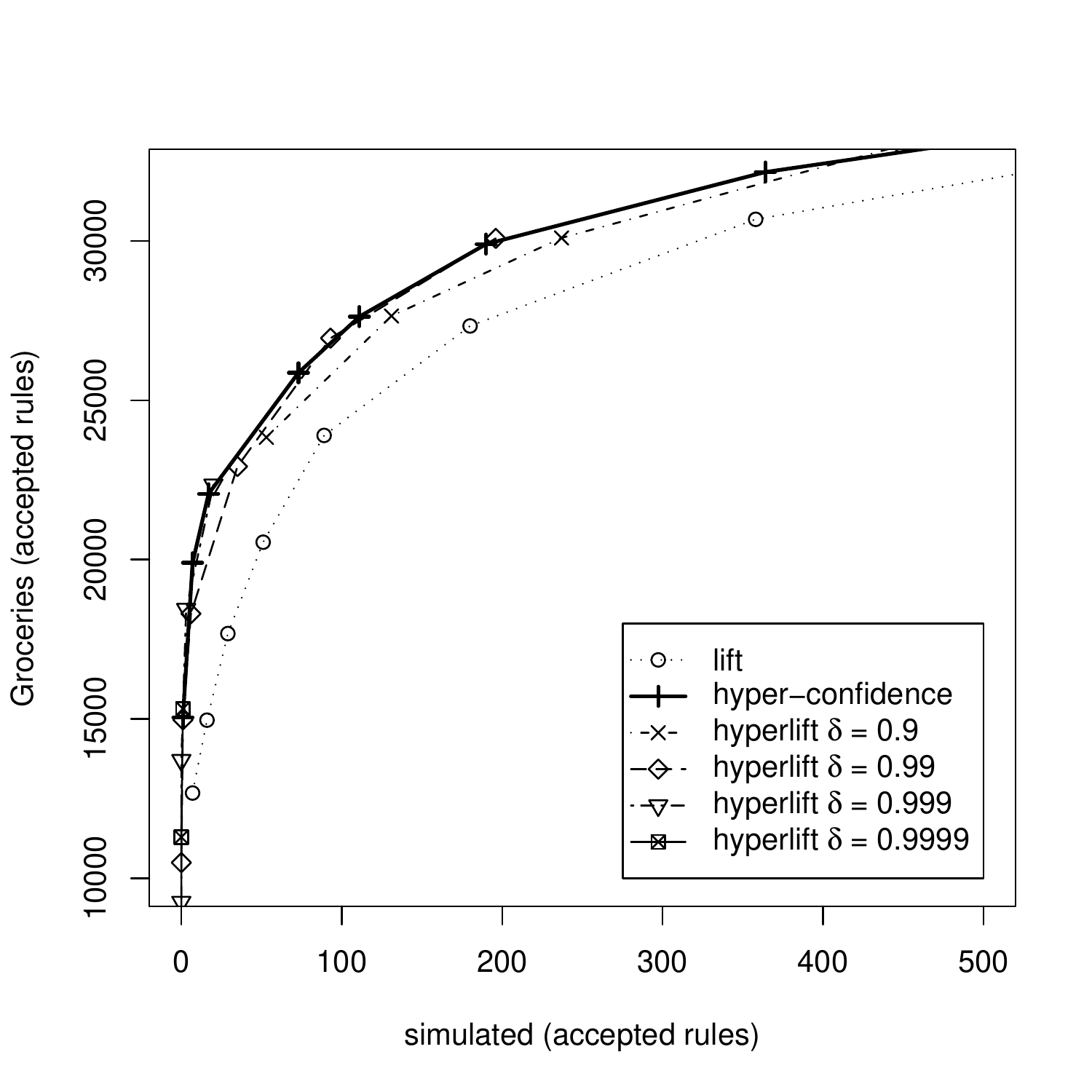}
\end{minipage}
\centerline{(a) Grocery}
\begin{minipage}[b]{.50\linewidth}
  \centering
  \includegraphics[width=\linewidth]{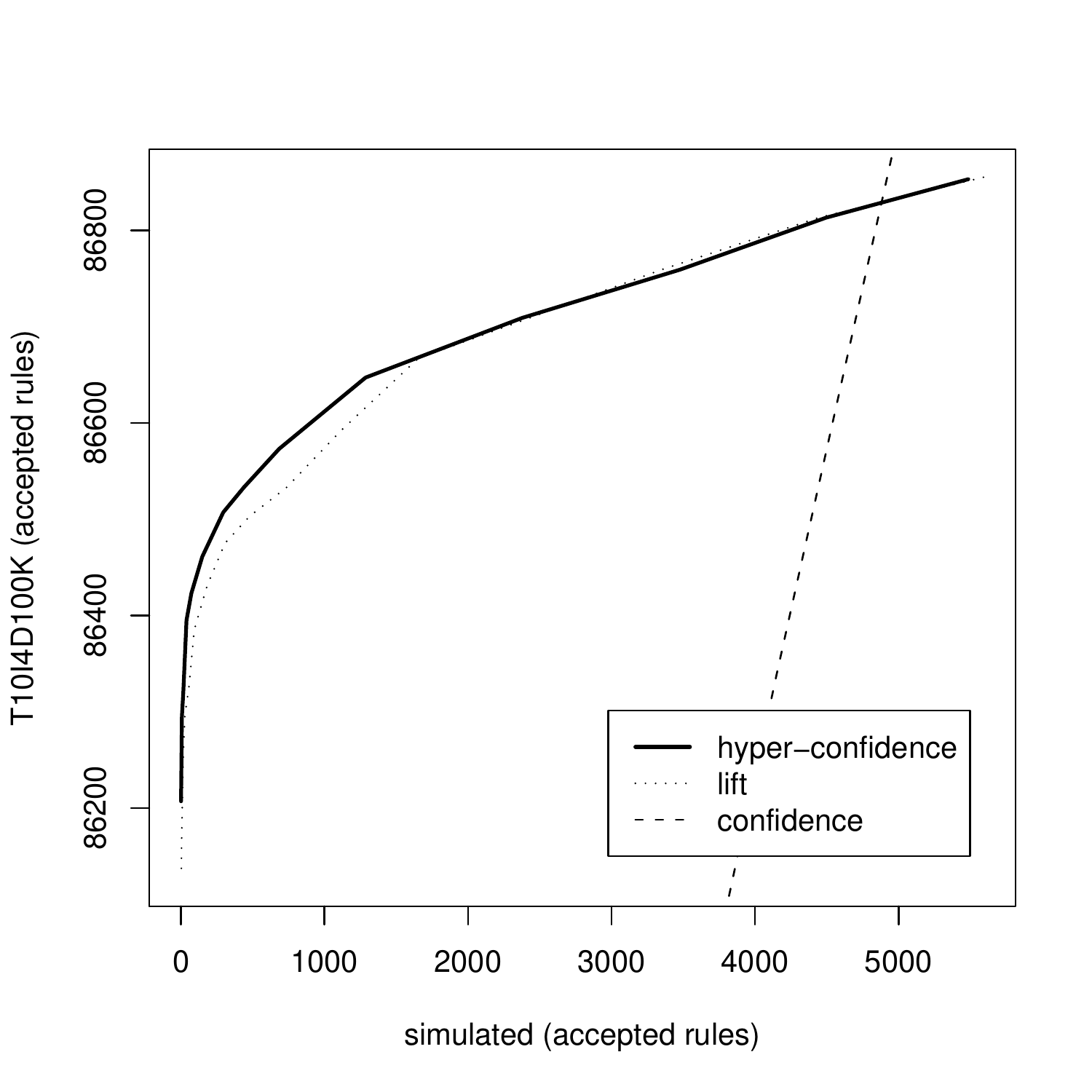}
\end{minipage}
  \begin{minipage}[b]{.50\linewidth}
    \centering
    \includegraphics[width=\linewidth]{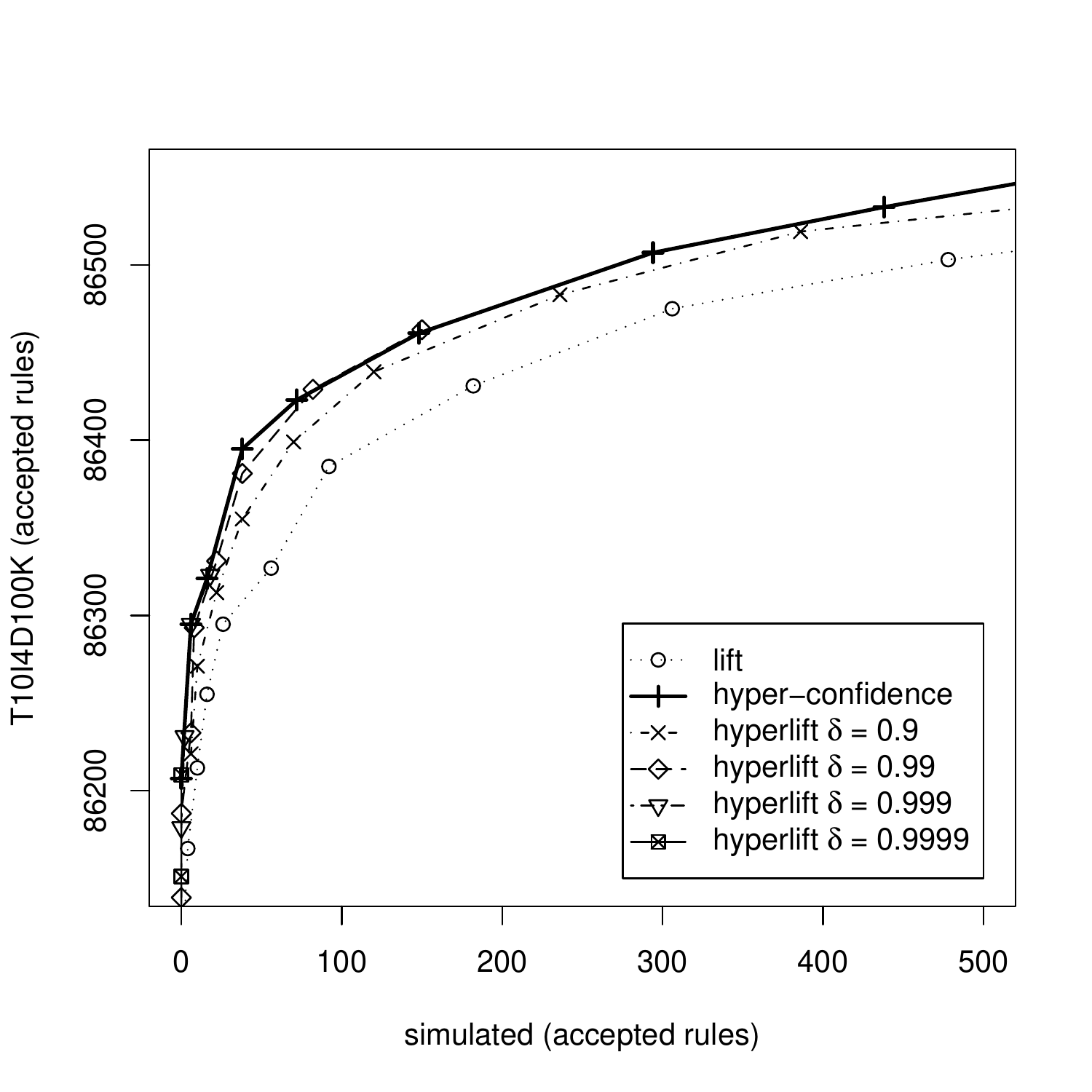}
\end{minipage}
\centerline{(b) T10I4D100K}
\begin{minipage}[b]{.50\linewidth}
  \centering
  \includegraphics[width=\linewidth]{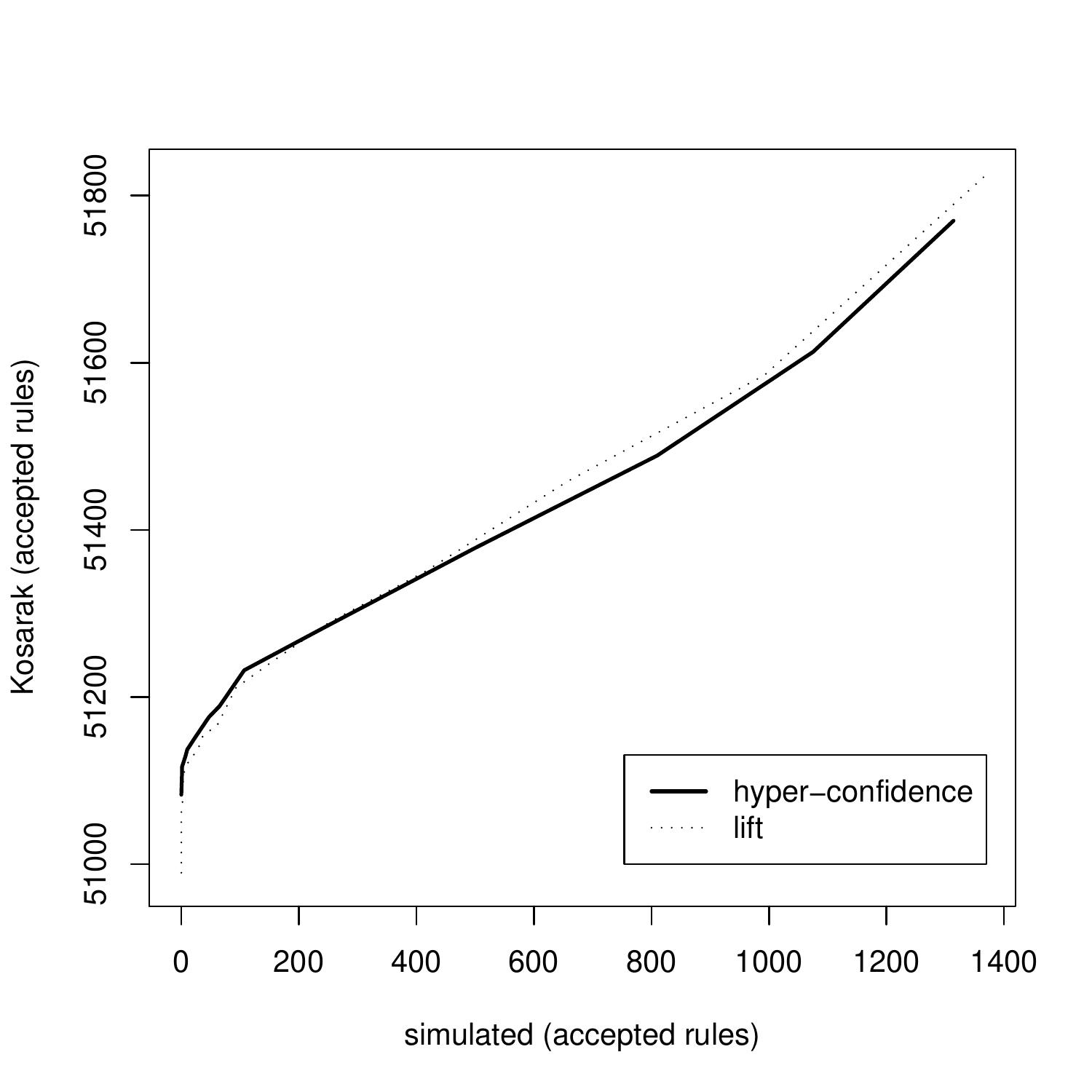}
\end{minipage}
  \begin{minipage}[b]{.50\linewidth}
    \centering
    \includegraphics[width=\linewidth]{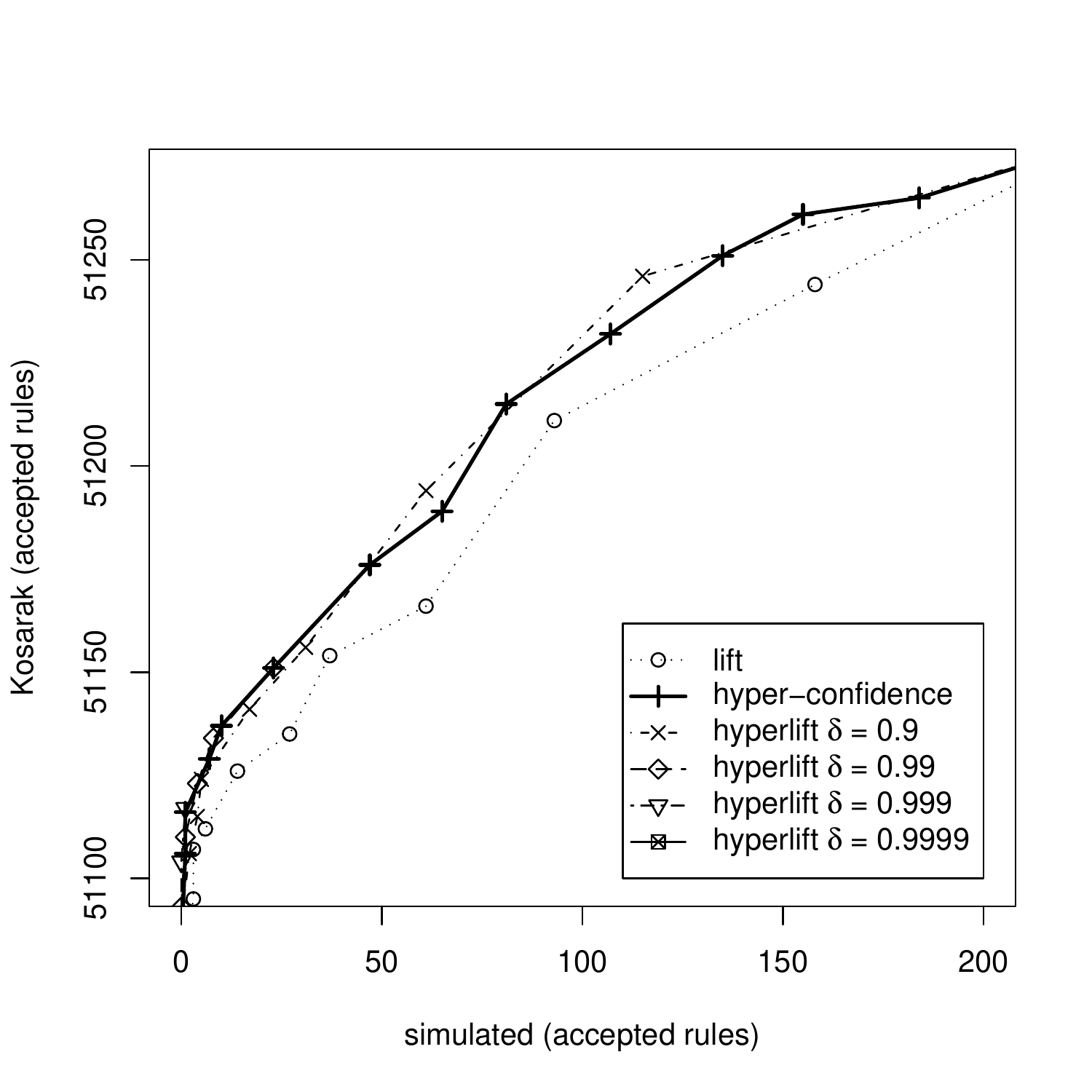}
\end{minipage}
\centerline{(c) Kosarak}
\caption{Comparison of number of rules accepted by different thresholds for 
lift, confidence, hyper-lift (only in the detail plots to the right) 
and hyper-confidence 
in the three databases and the simulated data sets.\label{fig:hyperlift_lift}}
\end{figure}

To analyze the trade-off in more detail, we proceed as follows:
We vary the threshold for lift (a minimum lift between 1 and 3) 
and assess the number of rules accepted in the databases and the
simulated data sets for each setting.
Then we repeat the procedure with confidence (a minimum between 0 and 1),
with hyper-lift (a minimum hyper-lift
between 1 and 3) at four settings for $\delta$ (0.9, 0.99, 0.999, 0.9999)
and with hyper-confidence (a minimum threshold between 0.5 and 0.9999).
We plot the number of accepted rules in the
real database by the number of accepted rules in the simulated data sets
where the points for each measure (lift, hyper-confidence, and 
hyper-lift with the same 
value for $\delta$) are connected by a line to form a curve.
The resulting plots in Figure~\ref{fig:hyperlift_lift} are 
similar in spirit to \emph{Receiver Operating Characteristic (ROC) plots} 
used in 
machine learning~\citep{misc:Provost+Fawcett:2001} to compare classifiers
and can be interpreted similarly.
Curves closer to the top left corner of the plot represent better results, 
since they provide a better ratio of true positives (here
potentially useful rules accepted in the real databases) and false positives 
(spurious rules accepted in the simulated data sets) regardless
of class or cost distributions. 

Confidence performs considerably worse than the other measures and is only
plotted in the left hand side plots.  For the Kosarak database, confidence
performs so badly that its curve lies even outside the plotting area. 

Over the whole range of parameter values presented in the left hand side
plots in Figure~\ref{fig:hyperlift_lift},
there is only little difference visible 
between lift and hyper-confidence visible.
The four hyper-lift curves are very close to the hyper-confidence curve and
are omitted from the plot for better visibility.
A closer inspection
of the range with few spurious rules accepted in the simulated data sets 
(right hand side plots in Figure~\ref{fig:hyperlift_lift}) 
shows that in this part hyper-confidence and hyper-lift clearly provides 
better results than lift (the new measures dominate lift).
The performance of hyper-confidence and hyper-lift are comparable.
The results for the Kosarak database look different than for the other two
databases. The reason for this is that the generation process of
click-stream data is very different from market basket data. For 
click-stream data the user clicks through a collection
of Web pages. On each page the hyperlink structure confines the user's choices 
to a usually very small subset of all pages. These restrictions are
not yet incorporated into the probabilistic framework. However, hyper-lift
and hyper-confidence do not depend on the framework and thus 
will produce still consistent results.

Note that in the previous evaluation, we did not know 
how many accepted rules in the real databases 
were spurious. However, we can speculate that if
the new measures suppress 
noise better for the simulated data, it also produces better results in the
real database and the improvement over lift  
is actually greater than can be seen in Figure~\ref{fig:hyperlift_lift}.

Only for synthetic data sets, where we can fully control the generation process,
we know which rules are non-spurious.
We modified the generator described by~\citet{arules:Agrawal1994} to
report all itemsets which were used in generating the data set.
These itemsets represent all non-spurious patterns 
contained in the data set.
The default parameters for the generator 
to produce the data set T10I4D100K
tend to produce 
easy to detect patterns since with the used so-called \emph{corruption level}
of 0.5 the 2000~patterns appear in the data set only slightly corrupted. 
We used a much higher corruption level of 0.9 which does not change the 
basic characteristics reported in Table~\ref{tab:datasets} above
but makes it considerably harder to find the non-spurious patterns.

\begin{figure}[pt]
  \centering
  \includegraphics[width=.8\linewidth]{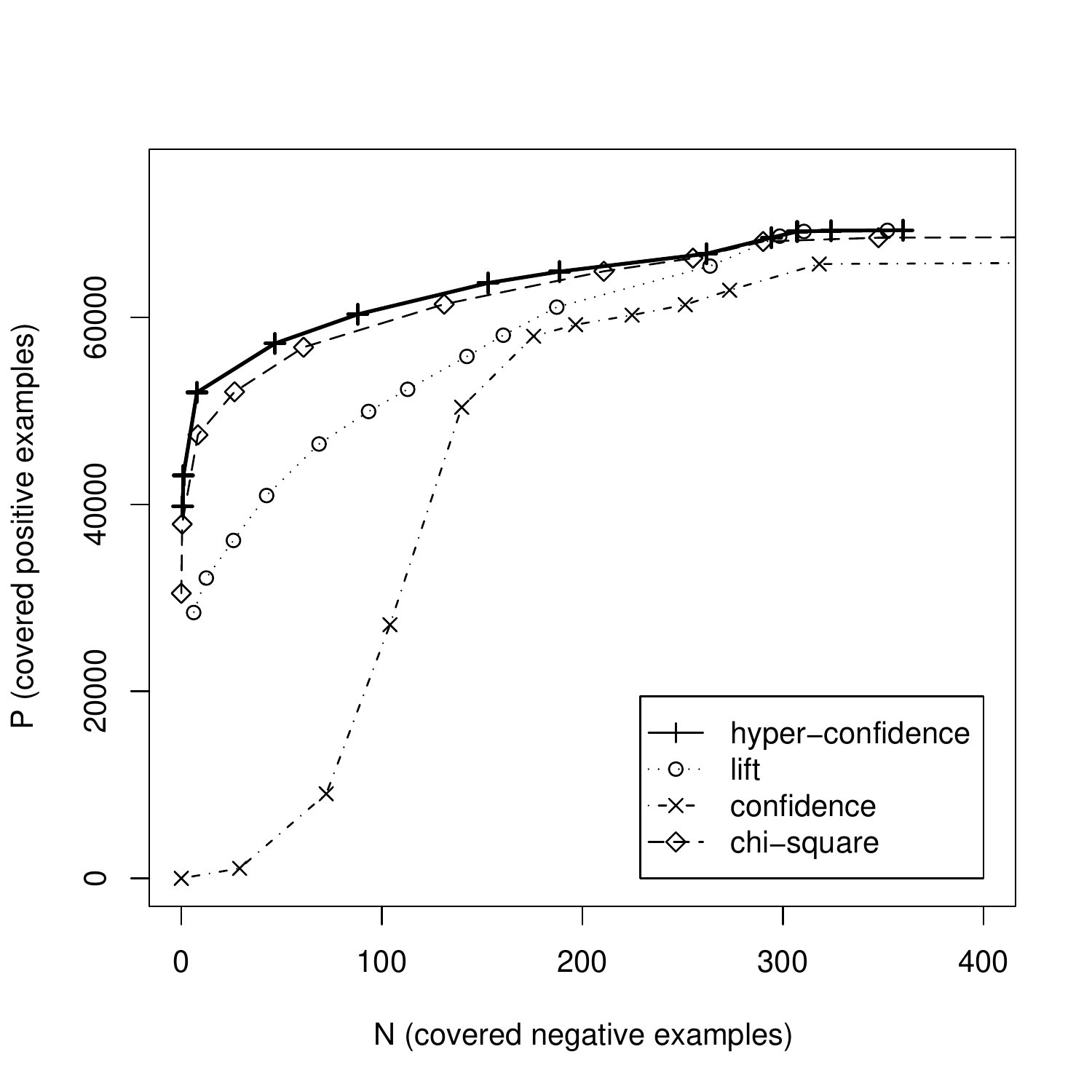}
  \caption{Average PN graph for 100 data sets generated with a 
  corruption rate of 0.9.\label{fig:hyperconf_lift}}
\end{figure}

We generated 100~data sets with 1000~items and 
100,000~transactions each, where we saved all 
patterns used for the generation. 
For each data set, we generate sets of rules which satisfy a 
minimum support of 0.001 and different thresholds for 
hyper-confidence, lift and confidence
(we omit hyper-lift here since the results 
are very close to hyper-confidence). 
For each set of rules, we count how many accepted rules represent
patterns which were used for generating the corresponding data set
(\emph{covered positive examples,~P}) and
how many rules are spurious (\emph{covered negative examples,~N}).
To compare the performance of the different measures in a single plot, 
we average the values for $P$ and $N$ for each measure at each used threshold
and plot the results (Figure~\ref{fig:hyperconf_lift}).

A plot of corresponding $P$ and $N$ values 
with all points for the same measure connected by a line
is called a 
\emph{PN graph} in the \emph{coverage space}
which is similar to the \emph{ROC space} without normalizing the 
X and Y-axes~\citep{misc:Fuernkranz+Flach:2005}. 
PN graphs can be interpreted similarly to ROC graphs: Points closer
to the top left corner indicate better performance.
Coverage space is used in this evaluation since, 
other than most classifiers, association rules typically 
only cover a small fraction of all examples 
(only rules generated from frequent itemsets generate rules) which makes
coverage space a more natural representation than ROC space.

Averaged PN graphs for hyper-confidence, lift, confidence
and the $\chi^2$ statistic are presented  
in Figure~\ref{fig:hyperconf_lift}.
Hyper-confidence dominates lift by a considerably larger margin than 
in the previous experiments reported in Figure~\ref{fig:hyperlift_lift}(b)
above. This supports the speculation that 
the improvements achievable with hyper-confidence are
also considerable for real world databases.
Using a varying
threshold on the $\chi^2$ statistic as proposed by \citet{arules:Liu1999}
performs better than lift and 
provides only slightly inferior results than hyper-confidence.

We also inspected the results for the individual data sets. While
the characteristics of the data sets vary sometimes significantly 
(due to the way the patterns used in the generation process are produced;
~see \cite{arules:Agrawal1994}),
all data sets show similar results with hyper-confidence dominating
all other measures.

\section{Conclusion\label{sec:conclusion}}

In this contribution we used a simple independence model (a null model with
``no structure'') to simulate a data set with comparable characteristics as a
real-world data set from a grocery outlet.  We visually compared the values of
different measures of interestingness for all possible rules with two items. In
the comparison we found the same problems for confidence and lift, which other
authors already pointed out. However, these authors only argued with specially
constructed and isolated example rules. The analysis used in this paper gives a
better picture of how strongly these problems influence the process of
selecting whole sets of rules.  Confidence favors rules with high-support items
in the right hand side of the rule. For databases with items with strongly
varying support counts, this effect dominates confidence which makes it a bad
measure for selecting or ranking rules. Lift has a strong tendency to produce
the highest values for rules which just pass the set minimum support threshold.
Selecting or ranking rules by lift will lead to very unstable results, since
even small changes of the minimum support threshold will lead to very different
rules being ranked highest.

Motivated by these problems, two novel measures of interestingness, hyper-lift
and hyper-confidence, are developed. Both measures quantify the deviation of
the data from a null model which models the co-occurrence count of two
independent itemsets in a database. 
Hyper-lift is similar to lift but uses
instead of the expected value a quantile from the corresponding hyper-geometric
distribution. The distribution can be very skewed and thus hyper-lift can
result in significantly different ordering of rules than lift. Hyper-confidence
is defined as the probability of realizing a count smaller than the observed
count and from its setup related to a one-sided Fisher's exact test.  

The new measures do not show the problematic behavior described for confidence
and lift above. Also, both measures outperform confidence, lift, and the
$\chi^2$ statistic on real-word data sets from different application domains as
well as in an experiment with simulated data. This indicates that 
the knowledge of how independent itemsets co-occur can be used to
construct superior measures of interestingness which improve the 
quality of the rule set returned by the mining algorithm.

A topic for future research is to develop more complicated independence 
models which incorporate constraints for specific application domains.
For example, in click-stream data, the link structure
restricts which pages can be reached from one page. 
Also the generation of artificial data sets
which incorporate models for dependencies between items is an important
area of research. Such data sets could greatly 
improve the way the effectiveness of data mining applications is evaluated
and compared.

\bibliographystyle{ida}
\bibliography{association_rules,recommender,marketing,stat,hahsler,R,misc}

\end{document}